\documentclass[preprint]{aastex}

\usepackage{amsfonts}
\usepackage{amsmath}

\shorttitle{LOCAL APPROXIMATIONS TO LARGE-SCALE STRUCTURE}
\shortauthors{Makler, Kodama and Calv\~{a}o}

\begin{document}

\title{ON LOCAL APPROXIMATIONS TO THE NONLINEAR EVOLUTION OF LARGE-SCALE STRUCTURE}

\author{Mart\'{\i}n Makler}
\affil{Centro Brasileiro de Pesquisas F\'{\i}sicas, Rua Xavier
Sigaud, 150\\ CEP 22290-180, Rio de Janeiro, RJ, Brazil\\
{\tt{martin@lafex.cbpf.br}}}

\and

\author{Takeshi Kodama and Maur\'{\i}cio
O.\ Calv\~{a}o} \affil{Universidade Federal do Rio de Janeiro,
Instituto de F\'{\i}sica, C. P. 68528\\ CEP 21945-970, Rio de
Janeiro, RJ, Brazil\\ {\tt{tkodama@if.ufrj.br, orca@if.ufrj.br}}}

\begin{abstract}

We present a comparative analysis of several methods, known as local
Lagrangian approximations, which are aimed to the description of the nonlinear 
evolution of large-scale structure. We have investigated various aspects of 
these approximations, such as the evolution of a homogeneous ellipsoid, 
collapse time as a function of initial conditions, and asymptotic behavior.
As one of the common features of the local approximations, we found that
the calculated collapse time decreases asymptotically with the inverse of the initial shear. Using these approximations, we have computed the cosmological mass function, finding reasonable agreement with $N$-body simulations and the Press-Schechter formula.

\end{abstract}


\section{INTRODUCTION}

Large-scale structures are believed to have formed from the gravitational
amplification of primordial perturbations. At its first stages, the process
of gravitational clustering can be investigated using linear perturbation
theory. However, as the universe evolves, nonlinear concentrations of mass
arise. Many structures we see today correspond to fluctuations several
orders of magnitude higher than the mean density of the universe; for
example, clusters of galaxies have typically $\rho _{\mathrm{cluster}}/\rho
_{\mathrm{universe}}\sim 10^{2}-10^{3}$. For larger scales this ratio
decreases, approaching unity in the largest structures.

As there is no analytical treatment for the nonlinear regime,
$N$-body simulations are often resorted to. The numerical
simulations had an enormous development in the last decade (see
Bertschinger 1998 and references therein), being able to
reproduce many features of the large scale structure. However they
do not always provide a clear insight of the physics of nonlinear
gravitational collapse. Moreover, they are usually very
time-consuming, making it difficult to scan a large part of the
parameter space of the cosmological models.

For this reason, semi-analytical methods have been devised to
tackle such a complex problem. The first approximation developed
to study the nonlinear regime was introduced by Zel'dovich (1970).
There are now various approximation schemes to analyze different
aspects of non-linear clustering, including extensions of the
Zel'dovich approximation (for a review see Sahni \& Coles 1995).
Among them, the so-called local Lagrangian approximations have
been introduced rather recently. The basic feature of these local
approximations is that the kinematical parameters in each fluid
element evolve independently of those of other elements. Thus the
time evolution of a self gravitating fluid is replaced by a set of
ordinary differential equations. This comes at the expense of
losing information about the positions of each fluid element. Only
local quantities, such as the density contrast, shear and
expansion rate can be determined.

Due to their handy applicability compared to the numerical
simulations, as seen in the case of the widely used Zel'dovich
approximation, they deserve a closer investigation. For some of
these methods, certain aspects of their performance and
applicability have already been discussed. However, to the
authors' knowledge, no systematic comparison among them has ever
been done. We consider it worthwhile to analyze them in a unified
way in order to exploit general properties of these
approximations, clarifying their similarities and differences. It
is also important to compare their performance in some practical
applications. In this paper, we discuss the following four
approximations, in addition to the Zel'dovich approximation: the
Local Tidal Approximation (Hui \& Bertschinger 1996), the
Deformation Tensor Approximation (Audit \& Alimi 1996), the
Complete Zel'dovich Approximation (Betancort-Rijo \&
L\'{o}pez-Corredoira 2000) and the Modified Zel'dovich
Approximation (Reisenegger \& Miralda-Escud\'{e} 1995). All of
them intend to be applicable to the highly non-linear regime. To
the best of our knowledge, these comprise all existent local
approximations in the literature, that are exact for planar,
spherical, and cylindrical symmetries (except for the Zel'dovich
approximation).

The paper is organized as follows: In section \ref{Aloc}, we
briefly review various local approximations in a unified way. In
section \ref{Applications}, these methods are applied to several
cases. First, we discuss the homogeneously collapsing ellipsoid.
We then study their behavior under general initial conditions.
Finally, we apply some of these approximations to the calculation
of the cosmological mass function. We sum up our results and
present conclusions in section \ref{conclusion}. In the appendices
we present useful formulae for the calculation of the mass
function together with fitting formulae for the collapse time in
the approximations considered here.

\section{LOCAL APPROXIMATIONS}

\label{Aloc}

Throughout this paper we will only consider the case of cold dark matter
(CDM), which is assumed to be collisionless, at least on large scales. This
is well justified since 80 to 90\% of the matter that clusters is composed
by CDM (Turner 2000, Durrer \& Novosyadlyj 2000). Furthermore, as long as
the trajectories do not intersect, we can treat the CDM as a pressureless
fluid.

We will be working in a matter-dominated flat universe (the
Einstein-de Sitter universe, hereafter EdS). Recent observational
evidences are consistent with a zero curvature universe (De
Bernardis et al. 2000; Hanany et al 2000). Even if we had a
non-flat universe we would only require that the curvature be
negligible in the scales of interest. The assumption of matter
dominance may seem unrealistic since the observations indicate
that the universe is now dominated by a repulsive homogeneous
cosmological term (Perlmuter et al. 1998; Riess et al. 1998;
Zehavi \& Dekel 1999). However, the energy density of this term
decays more slowly than the matter density.
In the case of a cosmological constant we would have $\rho _{\Lambda }=const$%
. whereas for matter we have $\rho _{M}\propto a^{-3}$, where $a$ is the
scale factor of the universe. Since most structures form at a time when $\rho_\Lambda \ll \rho_M$, we can safely ignore the effect of the 
cosmological term on the collapse process.

The peculiar motions in the universe are much smaller than the speed of
light. For perturbations on scales smaller than the Hubble radius, we can
use the Newtonian approximation to describe the gravitational clustering.
The basic equations for nonrelativistic pressureless matter in a perturbed
EdS universe are the Euler, the continuity and the Poisson equations
(Bertschinger 1996):
\begin{gather}
\frac{1}{a}\frac{dv_{i}}{d\tau }+\frac{\dot{a}}{a^{2}}v_{i}=-\frac{\partial
\phi }{\partial x^{i}},  \label{Euler} \\
\frac{d\delta }{d\tau }+a\left( 1+\delta \right) \theta =0,
\label{Continuity} \\
\frac{\partial ^{2}\phi }{\partial x^{i}\partial x_{i}}=4\pi Ga^{2}\bar{\rho}%
\delta ,  \label{Poisson}
\end{gather}
where $\delta =\left( \rho -\bar{\rho}\right) /\bar{%
\rho}$ is the density contrast $v_{i}=\left( dx_{i}/d\tau \right) /a$ is the
peculiar velocity, $\theta =\partial v^{i}/\partial x^{i}$ is the expansion,
$\phi $ is the peculiar gravitational potential, and the time variable $\tau
$ is related to the cosmic time $t$ (also known as proper time) by $d\tau =dt/a^{2}$. The comoving
coordinate $x_{i}$ is given in terms of the position $r_{i}$ by $%
x_{i}=r_{i}/a$. The left hand side of equation (\ref{Euler}) is simply $%
\left( d^{2}x_{i}/d\tau ^{2}\right) /a^{2}$, so that it looks like the usual
Euler equation (apart form the factor $a^{-2}$). In an Einstein-de Sitter
background the scale factor is proportional to $\tau ^{-2}$. We set $a=\tau
^{-2}$ such that $4\pi Ga^{2}\bar{\rho}=6\tau ^{2}=6/a$. The present value
of the scale factor $a_{0}$ is fixed to be unity.

The Lagrangian coordinates $q_{i}$ are often used instead of the position $%
x_{i}$ in nonlinear analyses. In terms of $q_{i}$ the convective derivative $%
d/d\tau =\partial /\partial \tau |_{x}+v_{i}\partial /\partial x_{i}$ is
simply given by the time derivative at fixed $q$: $d/d\tau =\partial
/\partial \tau |_{q}$. The Lagrangian coordinates are chosen to be the
initial comoving positions: $q_{i}=\lim_{a\rightarrow 0}r_{i}/a$.

The Jacobian matrix of the transformation $x_{i}\rightarrow q_{i}$%
\begin{equation}
J_{ij}=\frac{\partial x_{i}}{\partial q^{j}},
\end{equation}
is known as the deformation tensor. The velocity gradient $\partial
v_{i}/\partial x^{j}$ can be expressed in terms of $J_{ij}$ as
\begin{equation}
\frac{\partial v_{i}}{\partial x^{j}}=\frac{1}{a}J_{kj}^{-1}\frac{dJ_{i}^{k}%
}{d\tau }.  \label{vgrad}
\end{equation}
The density is given by $\rho \left( x,t\right) =\bar{\rho}J$, where $J$ is
the determinant of $J_{ij}$. It is easy to see that the continuity equation (%
\ref{Continuity}) is solved exactly with $\delta =J^{-1}-1$.

Differentiating equation (\ref{Euler}) with respect to $x_{j}$ we find
\begin{equation}
J_{jk}^{-1}\frac{d^{2}J_{i}^{k}}{d\tau ^{2}}\frac{1}{a^{2}}=-\frac{\partial
\phi }{\partial x^{i}\partial x^{j}},  \label{divPhi}
\end{equation}
whose trace furnishes Raychaudhuri's equation
\begin{equation}
J_{ij}^{-1}\frac{d^{2}J^{ji}}{d\tau ^{2}}=-4\pi Ga^{4}\bar{\rho}\left(
J^{-1}-1\right) .  \label{Raychaudhuri}
\end{equation}
This is a local equation for $J_{ij}$ in the sense that it has no spatial
derivatives, although it is not sufficient for determining the nine
components of the deformation tensor. Usually this equation is written in
terms of the kinematical parameters, $\theta ,$ $\sigma _{ij}$ (shear) and $%
\omega _{i}$ (vorticity), defined by:

\begin{equation}
\frac{\partial v_{j}}{\partial x_{i}}={\frac{1}{3}}\,\theta \,\delta
_{ij}+\sigma _{ij}+\omega _{ij}\,\quad \sigma _{ij}=\sigma _{ji}\ ,\quad
\omega _{ij}=\epsilon _{ijk}\,\omega ^{k}=-\omega _{ji}.  \label{gradv}
\end{equation}
If the initial conditions have no vorticity, then we have $\omega
^{i}=0$ during all the evolution, as long as the trajectories do not intersect. Here we will consider only the case of vanishing vorticity.

Equations (\ref{Euler}) to (\ref{Poisson}) form a set of nonlinear partial
differential equations. However, for certain specific configurations the
time evolution of the deformation tensor $J_{ij}$ behaves as if each space
point evolves independently from the others. One might then expect that for
more general situations the locality may hold, at least approximately, for
these variables. Accordingly, several methods have been introduced which are
known as local approximations. In their framework, the influence of the
neighbors may enter only through the initial conditions.

In addition to the solution of the continuity and Euler equations,
the local approximations discussed here will replace the
essentially nonlocal exact equation (\ref{Poisson}) either by some
Ansatz inspired on equation (\ref {Raychaudhuri}), or by local
evolution equations for the second derivative of the peculiar
gravitational potential $\phi $ (see also Kofman \& Pogosian
1995 for a discussion).

One of the basic features of local approximations is that the
eigenvectors of the deformation tensor do not change with time. Thus, once diagonalized, $J_{ij}$ remains diagonal in the same frame, along all the
evolution. This condition is either assumed from the beginning or
appears as a consequence of the approximation introduced in the
evolution equations. Actually, this
assumption is not strictly consistent with the evolution of the mapping $%
q_{i}\rightarrow x_{i}$, so that the reconstruction of space coordinates in
these local approximations is not possible (cf. subsection \ref{features}).

In the basis where $J_{ij}$ is diagonal,
\begin{equation}
J_{ij}=\left( 1+w_{i}\right) \delta _{ij},
\end{equation}
Raychaudhuri's equation (\ref{Raychaudhuri}) is written as
\begin{equation}
\sum_{i=1}^{3}\frac{\ddot{w}_{i}}{\left( 1+w_{i}\right) }=-4\pi Ga^{4}\bar{%
\rho}\left( \frac{1}{\left( 1+w_{1}\right) \left( 1+w_{2}\right) \left(
1+w_{3}\right) }-1\right) .  \label{EM}
\end{equation}
The local approximations discussed here are required to be exact for planar,
spherical and cylindrical symmetries. In the spherical case we have $%
w_{1}=w_{2}=w_{3}$; for a cylindrical perturbation $w_{1}=w_{2}$, and $%
w_{3}=0;$ whereas for planar symmetry $w_{2}=w_{3}=0$. In these three cases,
as we have only one independent eigenvalue $w_{i}$ of the deformation
tensor, this equation can be solved for $w_{i}$.

\subsection{Zel'dovich Approximation}

The \textit{Zel'dovich approximation} (Zel'dovich 1970), hereafter ZA, can
be viewed as a solution of the linearized form of equation (\ref{EM}):
\begin{equation}
\sum_{i=1}^{3}\ddot{w}_{i}=4\pi Ga^{4}\bar{\rho}\sum_{i=1}^{3}w_{i}.
\label{Lin}
\end{equation}
Zel'dovich used the solution of the linearized equations
(\ref{Euler} to \ref {Poisson}) $x_{i}=q_{i}-D\left( t\right) \Psi
_{i}\left( q\right) $ and extrapolated it into the nonlinear
regime. The eigenvalues of the deformation tensor are thus given by
\begin{equation}
w_{i}=-D\left( t\right) \lambda _{i}^{0}\left( q\right) ,  \label{ZA}
\end{equation}
where $\lambda _{i}^{0}$ are the eigenvalues of$\ \partial \Psi
_{i}/\partial x^{j}$. Substituting this expression into the eq. (\ref
{Lin}) we find two solutions for $D$, known as the growing and decaying
modes. For an EdS universe we have
\begin{equation}
w_{i}^{+}=-^{+}\lambda _{i}^{0}a\text{\quad and\quad }w_{i}^{-}=-^{-}\lambda
_{i}^{0}a^{-3/2}.  \label{wlin}
\end{equation}
Since the decaying mode becomes negligible very quickly, only the growing
mode is relevant for our discussion. The initial conditions are specified in
terms of the $\lambda _{i}^{0}$, which are functions of the initial
positions $q$. The principal axes of $\partial \Psi _{i}/\partial x^{j}$ are
generally different for each point. We will denote the linear growing mode
solution by $-\lambda _{i}$:
\begin{equation}
\lambda _{i}=\lambda _{i}^{0}a.  \label{lambdai}
\end{equation}
In the linear regime the density contrast $\delta $ will be given by $\delta
_{L}=\left( \lambda _{1}^{0}+\lambda _{2}^{0}+\lambda _{3}^{0}\right)
a=\delta _{0}a$.

The gist of the Zel'dovich approximation is that the linearized
trajectories can lead to nonlinear density perturbations.
Analogous ideas have been applied in many approximations. An
example is the higher-order Lagrangian expansions, where the
perturbed quantity is the displacement
field. In an EdS universe the solution may be written in the form $%
x_{i}=q_{i}+\sum_{n=1}a^{n}\Psi _{i}^{(n)}\left( q\right) .$ The
first order solution $\Psi _{i}^{(1)}$ is the Zel'dovich
approximation. The determination of the higher order $\Psi
_{i}^{(n)}$ follows from the lower order ones through the solution
of Poisson equations. The second order solution is known as
Post-Zel'dovich approximation (Moutarde at al. 1991; Buchert 1992;
Lachi\`{e}ze-Rey 1993), and the third order is called
Post-post-Zel'dovich (Juszkiewicz, Bouchet, \& Colombi 1993; Buchert 1994).

The Zel'dovich approximation is widely used for the weakly non-linear
regime, and for generating initial conditions for numerical simulations. It
gives the exact solution for the case of planar symmetry.

\subsection{Modified Zel'dovich Approximation}

In the ZA the time factor in eq. (\ref{ZA}) is independent of the initial
conditions, and it is valid only for the linearized limit in $w_{i}$ (eq.
\ref{Lin}). Reisenegger \& Miralda-Escud\'{e} (1995) have proposed a
generalization of the Zel'dovich approximation where $D$ may depend on the
position through the initial conditions $\lambda _{i}^{0}$. The Ansatz $%
w_{i}=-D\left( \tau,\lambda^0_i\right) \lambda _{i}^{0}\left(
q\right) $ is substituted in equation (\ref{EM}) to give
\begin{equation}
\frac{d^{2}D}{d\tau ^{2}}=4\pi G\bar{\rho}a^{4}\frac{\eta _{1}D-\eta
_{2}D^{2}+\eta _{3}D^{3}}{\eta _{1}-2\eta _{2}D+3\eta _{3}D^{2}},
\label{MZA}
\end{equation}
where $\eta _{1}=\lambda _{1}^{0}+\lambda _{2}^{0}+\lambda
_{3}^{0}$, $\eta _{2}=\lambda _{1}^{0}\lambda _{2}^{0}+\lambda
_{1}^{0}\lambda _{3}^{0}+\lambda _{2}^{0}\lambda _{3}^{0}$ e $\eta
_{3}=\lambda _{1}^{0}\lambda _{2}^{0}\lambda _{3}^{0}$. This
equation, which must be solved numerically, determines completely
the function $D(\tau,\lambda _{i}^{0})$ and defines the
\textit{Modified Zel'dovich Approximation,} hereafter MZA. It is
exact for spherical, planar and cylindrical symmetries. However,
for underdense regions ($\delta _{0}<0$), the MZA may not work, as
pointed out by Reisenegger \& Miralda-Escud\'{e} (1995). This is
due to the fact that, when not all the three eigenvalues $\lambda
_{i}$ have the same sign, the denominator in the right hand side
of eq. (\ref{MZA}) will eventually vanish. Thus MZA cannot be used
with this kind of initial conditions.

\subsection{Deformation Tensor Approximation}

In the two local approximations discussed above, the time dependence of the
three $w_i$ is the same and it can be completely determined from equation
(\ref{EM}). The next two approximations will provide an equation for each of
the three $w_{i}$ and an analytical solution of $w_{i}$ in terms of the
linear solution $\lambda _{i}$ (eq. \ref{lambdai}). Due to the symmetry
among the axes, both the equation for $w_{i}$ and the explicit solution in
terms of $\lambda _{i}$ should be invariant under any exchange of indices, $%
\left( i,j,k\right) $.

Equation (\ref{EM}) may be written in the form
\begin{equation}
\sum_{i=1}^{3}\left[ \left( 1+w_{j}+w_{k}+w_{j}w_{k}\right) \ddot{w}%
_{i}-4\pi Ga^{4}\bar{\rho}\left( 1+\frac{w_{j}+w_{k}}{2}+\frac{w_{j}w_{k}}{3}%
\right) w_{i}\right]=0 ,  \label{Poisson2}
\end{equation}
where $\left( i,j,k\right) $ is a permutation of $\left(
1,2,3\right) $. Audit \& Alimi (1996) have, as an Ansatz, split
this equation into three equations for each $w_{i}$:

\begin{equation}
\left( 1+w_{j}+w_{k}+w_{j}w_{k}\right) \ddot{w}_{i}=4\pi Ga^{4}\bar{\rho}%
\left( 1+\frac{w_{j}+w_{k}}{2}+\frac{w_{j}w_{k}}{3}\right) w_{i}\text{.}
\label{DTA}
\end{equation}
This equation defines the \textit{deformation tensor approximation,}
hereafter DTA. Another motivation for the above equation is that it is exact
for planar, spherical and cylindrical perturbations. We have thus a set of
local equations that allows to determine each $w_{i}$ completely. Of course
this spliting of equation (\ref{Poisson2}) is not unique and we could add
more local terms in equation (\ref{DTA}) which would obey the symmetry
requirement.

\subsection{Complete Zel'dovich Approximation}

The \textit{Complete Zel'dovich Approximation,} CZA (Betancort-Rijo \&
L\'{o}pez-Corredoira 2000) assumes that the $w_{i}$ can be expanded in terms
of the linear solution $\lambda _{i}$ (eq. \ref{lambdai}). To satisfy the
symmetries required, the power series must have the following expression:

\begin{equation}
r_{i}(\lambda _{i},\lambda _{j},\lambda _{k})=1+\sum_{l,m,n=0}^{\infty
}C_{l,m,n}^{p}(\lambda _{j}+\lambda _{k})^{l}(\lambda _{j}-\lambda
_{k})^{2n}\lambda _{i}^{m},  \label{CZA}
\end{equation}
where
\begin{equation}
w_{i}=-\lambda _{i}r_{i},
\end{equation}
and $C_{l,m,n}^{p}$ are the coefficients of the $p$-th order terms, with $%
p\equiv l+2n+m$. The Zel'dovich approximation corresponds to $r_{i}=1$. The
second order term
\begin{equation}
w_{i}^{(2)}=-\lambda _{i}\frac{3}{14}\left( \lambda _{j}+\lambda _{k}\right)
\label{DTA2}
\end{equation}
coincides with that of the DTA.

For planar configurations one should have $r_{i}=1,$ thus
$C_{0,m,0}^{p}=0$. The other coefficients of the expansion are
determined from equations (\ref {Euler}) and (\ref{Poisson})
through a recursive scheme. Betancort-Rijo \& L\'{o}pez-Corredoira
(2000) calculated explicitly the coefficients $C_{l,m,n}^{p}$ up
to the terms of fourth order in $\lambda $ in an EdS universe.

When the higher-order terms become important, all of them
contribute roughly the same. Thus Betancort-Rijo \&
L\'{o}pez-Corredoira have chosen to truncate the series at the
fourth order and approximate the rest by a function $R(\lambda
_{i},\lambda _{j},\lambda _{k})$. This function is parametrized in
such a way that the result is in agreement with the exact planar,
spherical and cylindrical dynamics. Their expression for $R$ is:
\begin{eqnarray}
R(\lambda _{i},\lambda _{j},\lambda _{k}) & = &\left[ 1-9\left( \lambda _{i}-%
\frac{\lambda _{j}+\lambda _{k}}{2}\right) \left( 1-\frac{\lambda
_{i}+\lambda _{j}+\lambda _{k}}{1.3}\right) \right] \nonumber \\
 & & \left( R_{\mathrm{sp}}(\lambda _{i}+\lambda _{j}+\lambda _{k})-R_{%
\mathrm{sp}}(\lambda _{i})+R_{\mathrm{sp}}\left( \frac{\lambda _{j}+\lambda
_{k}}{2}\right) \right) ,  \label{R}
\end{eqnarray}
where $R_{\mathrm{sp}}$ is the correction term $R$, corresponding to the
spherical symmetry. By comparing the numerical results for overdense
perturbations with the truncated series solution, they fitted $R_{\mathrm{sp}%
}$ as
\begin{equation}
R_{\mathrm{sp}}\left( x\right) =2.58\times 10^{-3}x^{5}\left( 1-\frac{x}{2.06%
}\right) ^{-1}.
\end{equation}

The expansion (\ref{CZA}) up to $p=4$, together with expression
(\ref{R}), gives nearly exact results for spherical and
cylindrical overdense perturbations, and the exact result in the
planar case. Indeed, the CZA predicts that a spherical perturbation with 
$\delta_0=1$ will collapse at $a_c=1.72$, whereas the exact solution 
gives $a_c=1.69$.

The CZA does not apply for perturbations with negative values of $\lambda
_{i}^{0}$. For example, when all the three $\lambda _{i}^{0}$ are negative
the volume element should expand indefinitely, hence the $\lambda _{i}$ will
approach infinity and therefore the series expansion breaks down. It can be
easily seen that, as all $C_{l,m,n}^{p}$ are positive, if we truncate the
series in an odd power of $\lambda ,$ $r$ will change sign, and the fluid
element will eventually collapse.

\subsection{Local Tidal Approximation}

The general relativistic equations for the kinematical parameters in the
projection formalism are very akin to the Newtonian ones (see Ellis 1973).
The analog of the Poisson equation is obtained from the equation for the
Weyl tensor. Barnes \& Rowlinson (1989) pointed out that by neglecting the
magnetic part of Weyl tensor $H_{\mu \nu }$, the evolution
equation for the electric part $E_{\mu \nu }$ becomes local. 
The dynamics of kinematical
parameters is then reduced to a closed set of local equations. This result was
first applied to structure formation by Matarrese, Pantano, \& Saez (1993).
Since the magnetic part of the Weyl tensor has no Newtonian analogue,
Bertschinger \& Jain (1994) introduced the non-magnetic approximation, by
simply discarding the magnetic part $H_{\mu \nu }$ in the equation for 
$E_{\mu \nu }$ in the application to
the Newtonian cosmology. This approximation is exact for spherical and
planar configurations, but fails for cylindrical symmetry. Also it was not
able to reproduce the dynamics of the collapse even for a homogeneous
ellipsoid. Thus we will not consider this approximation further in this work.

Bertschinger \& Hamilton (1994) pointed out that, in a
``Newtonian'' limit, the role of the magnetic part $H_{ij}$ is not
altogether negligible (see also Ellis \& Dunsby 1997). Within this
framework, Hui \& Bertschinger (1996) have proposed the
\textit{local tidal approximation} (LTA), which consists in
discarding some terms in the evolution equation for $E_{ij}$, to
get

\begin{equation}
\frac{dE_{ij}}{d\tau }+\frac{1}{a}\frac{da}{d\tau }E_{ij}=-4\pi Ga^{3}\bar{%
\rho}\sigma _{ij},  \label{LTA}
\end{equation}
\bigskip where $E_{ij}$ is the Newtonian limit of $E_{\mu\nu}$, which gives the tidal field:
\begin{equation}
E_{ij}=\frac{\partial ^{2}\phi }{\partial x^{i}\partial x^{j}}-\frac{1}{3}%
\frac{\partial ^{2}\phi }{\partial x^{k}\partial x_{k}}\delta _{ij}=\frac{%
\partial ^{2}\phi }{\partial x^{i}\partial x^{j}}-\frac{4\pi Ga^{2}\bar{\rho}%
\delta }{3}\delta _{ij}.  \label{Eij}
\end{equation}
Equations (\ref{LTA}) and (\ref{divPhi}) written in terms of $E_{ij}$ form a
closed set of local equations. It was shown that the LTA is exact for
spherical, planar and cylindrical symmetries. In general, it is exact
whenever the orientation and axis ratios of the gravitational and velocity equipotentials are equal and constant for the mass element under consideration (Hui \& Bertschinger 1996).

It is possible to show that in the LTA, once the velocity gradient is
diagonalized, it will remain diagonal (Hui \& Bertschinger 1996) and so will
the deformation tensor. We may write equation (\ref{LTA}) in terms of $w_{i}$
by using the equation (\ref{Eij}) together with equation (\ref{divPhi}). 
We will then have a
set of three third-order equations for $w_{i}$ which completely determines
their evolution, once appropriate initial conditions are provided.
Alternatively, equation (\ref{LTA}) can be solved in terms of the
kinematical parameters. In this case, the $w_{i}$ are calculated by 
(eqs. (\ref{vgrad}) and (\ref{gradv}))

\begin{equation}
\frac{dw_{i}}{d\tau }=a\left( \sigma _{i}+\frac{1}{3}\theta \right) \left(
1+w_{i}\right) ,
\end{equation}
where $\sigma _{i}$ are the eigenvalues of the shear $\sigma _{ij}$.

\subsection{General Features\label{features}}

The local approximations discussed above are either a system of ordinary
differential equations or explicit expressions in terms of the linear
solution. In these approximations each point evolves independently of the others. The
influence of the other fluid elements enters only through the initial
conditions. They give the time evolution of the deformation tensor, and thus
the kinematical parameters for each volume element.

These local approximations are exact under some geometrical symmetries. In
particular, they are exact whenever $w_{1}=w_{2}=w_{3}$; or $w_{1}=w_{2}$,
with $w_{3}=0;$ or $w_{2}=w_{3}=0$ are satisfied locally. They are
nonperturbative, i.e., valid, in principle, for any $\delta $ or $\lambda
_{i}$.

The second order solution of the CZA and DTA (eq. \ref{DTA2}) is in agreement with the result of second
order Lagrangian perturbation theory for the density contrast (Sahni \& Coles 1995, Betancort-Rijo \& L\'{o}pez-Corredoira 2000). The ZA fails at second order.

The local approximations are not appropriate for recovering the positions.
To see this, let us consider an initial configuration such that the
deformation tensor is diagonal at every point
\begin{equation}
J_{ij}=\frac{\partial x_{i}}{\partial q^{j}}=\left( 1+w_{i}\right) \delta
_{ij}.  \label{diagonal}
\end{equation}
If this holds initially, it will be valid throughout the evolution, according
to the local approximations. In this case $x_{i}$ would be given by
\begin{equation}
x_{i}=\int \left( 1+w_{i}\right) dq_{i}.
\end{equation}
If $w_{i}$ had an explicit dependence on $q_{j}$ or $q_{k},$ nondiagonal
terms would arise in equation (\ref{diagonal}); hence $w_{i}$ must be a 
function of $q_{i}$ only.
Consequently, for this particular choice of initial conditions, each $%
\lambda _{i}^{0}$ must depend only on the coordinate $q_{i}$. However, as the $%
w_{i}$ evolve, they will in general depend on the three $\lambda _{i}^{0}$
and, ultimately, on the three coordinates: $w_{i}=w_{i}\left( \lambda
_{1}^{0}\left( q_{1}\right) ,\lambda _{2}^{0}\left( q_{2}\right) ,\lambda
_{3}^{0}\left( q_{3}\right) ,\tau \right) $. Thus equation (\ref{diagonal})
can no longer be satisfied. This shows that the local approximations in
general violate the integrability of the deformation tensor. In other words,
we cannot recover the actual positions in the local approximations (except
when $w_{i}$ is independent of the initial position). 

Another way of seeing that the integrability is violated is as follows:
If it were possible to reconstruct $x_1$ from $J_{1j}$, for example, then 
$J_{1j}$ ought to be a gradient field in lagrangian space. Therefore its 
curl should vanish. In the particular case of equation (\ref{diagonal}),
this implies that $\partial w_{1}/\partial q^{2}=0$ and 
$\partial w_{1}/\partial q^{3}=0$. These conditions are satisfied, in general, only by the ZA
for which $w_{i}=-a\lambda _{i}^{0}$ (provided that 
$\lambda _{i}^{0}=\lambda _{i}^{0}\left(q_{i}\right)$). Thus, the only approximation
which always permits the direct computation of the positions is the
Zel'dovich approximation. In spite of the non-integrability, these methods offer an approximate solution for the deformation tensor, allowing to calculate local
quantities, such as the kinematical parameters.

If any eigenvalues of the deformation tensor approach $-1$, the
density contrast $\delta$ will diverge. Since they are functions only
of $a$ and the initial conditions, $\lambda
_{i}^{0}$,  we can expand them near the collapse time $%
a_{c} $, as
\begin{equation}
w_{i}=-1-\left. \frac{dw_{i}}{da}\right| _{a=a_{c}}\left(
a_{c}-a\right) +\cdots .
\end{equation}
Therefore, the density contrast $%
\delta $ behaves, at the collapse time, as
\begin{equation}
\delta \propto \left( a_{c}-a\right) ^{-\gamma },
\end{equation}
where $\gamma $ is the dimensionality of the collapse ($\gamma =1$
for the collapse in only one axis, $\gamma =2$ for the collapse in
two axes simultaneously, and $\gamma =3$ for the collapse in three
axes). On the other hand, for the expansion $\theta $, we have from
equation (\ref{Continuity})  
\begin{equation}
\theta \rightarrow -\frac{2\sqrt{a_c}\gamma }{\left(
a_{c}-a\right) }
\end{equation}
for $a\rightarrow a_{c}$. The asymptotic behavior of the other kinematical
parameters can also be determined in a similar fashion.

\section{APPLICATIONS}

\label{Applications}

In order to compare the performance of these approximation schemes, we apply
them to some specific situations in the following subsections.

\subsection{The homogeneous ellipsoid}

\label{ellipsoid}

An initially homogeneous ellipsoid in an expanding universe develops in such
a way that the homogeneity is almost preserved during all the evolution.
Therefore, the \textit{homogeneously collapsing ellipsoid} model (HCE) is
considered to be very accurate (Eisenstein \& Loeb 1995; Hui \& Bertschinger
1996). The result of local approximations have been compared to this model.
Such a comparison is useful since it offers the possibility of checking
these approximations in a less symmetrical situation (Hui \& Bertschinger
1996; Audit \& Alimi 1996; Betancort-Rijo \& L\'{o}pez-Corredoira 2000). It
is worthwhile to compare theses analyses including the MZA.

The equation of motion for the HCE is given by (Icke 1973; White \& Silk
1979)
\begin{equation}
\frac{d^{2}Y_{i}}{d\tau ^{2}}=-\frac{2}{9}aY_{i}\left(
X_{1}X_{2}X_{3}-Y_{1}Y_{2}Y_{3}\right) C_{D}(Y_{k}^{2},Y_{j}^{2},Y_{i}^{2}),
\label{ellipEq}
\end{equation}
where, as before, $\left( i,j,k\right) $ are permutations of
$\left( 1,2,3\right) $, $Y_{i}$ represent the axes of the
ellipsoid in comoving coordinates, and $X_{i}$ are their
asymptotic values for $a\rightarrow 0$. The function $C_{D}$ is
the degenerate case of Carlson's integral of the third kind
(Carlson 1977; Press et al. 1992):
\begin{equation*}
C_{D}(x,y,z)=\frac{3}{2}\int_{0}^{\infty }\frac{ds}{\left( z+s\right) ^{3/2}%
\sqrt{\left( x+s\right) \left( y+s\right) }}.
\end{equation*}
The linear growing mode is
\begin{equation}
Y_{i}\underset{a\rightarrow 0}{\simeq }X_{i}\left( 1-\frac{1}{3}%
X_{1}X_{2}X_{3}\,C_{D}(X_{k}^{2},X_{j}^{2},X_{i}^{2})\delta _{0}a\right) .
\end{equation}

As there is no rotation, the orientation of the principal axes
does not change. Thus, the position of each element will be
proportional to the expansion in each direction. If an element inside
the ellipsoid has initial position $q_{i}$, then its coordinates
at a later time will be given by
\begin{equation}
x_{i}=\frac{Y_{i}}{X_{i}}q_{i}.  \label{relips}
\end{equation}
With this expression we may compute the kinematical parameters which will
not depend on the position. The same holds for the tidal field. Hence we can
compute the evolution of a fluid element according to the local
approximations, and compare\ with the evolution of $E_{ij}$, $\sigma _{ij}$,
$\delta $ and $\theta $ as derived from the ellipsoid solution with the same
initial conditions.

From expression (\ref{relips}) we can see that the deformation tensor does
not depend on $q_{i}$ inside the ellipsoid:
\begin{equation}
J_{ij}=\frac{Y_{i}}{X_{i}}\delta _{ij}.
\end{equation}

As discussed in the previous section, the positions of the fluid elements
may not necessarily be recoverable in the local approximations. However, the
choice of the same $w_{i}$ for any fluid element ($w_{i}$ independent of $%
q_{i}$) is consistent with the HCE. In this case, we may recover the
positions from the $w_{i}$ as
\begin{equation}
x_{i}=\left( 1+w_{i}\right) q_{i}.
\end{equation}

\begin{figure}
\plotone{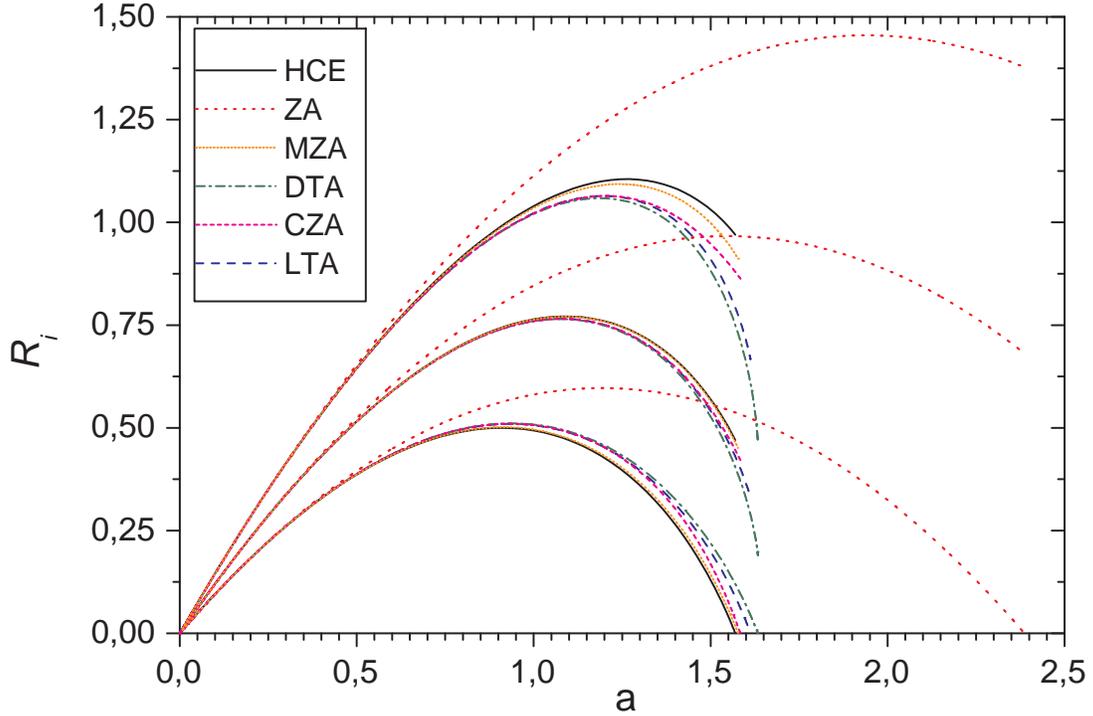} 
\caption{Evolution of the three axes $R_i$ of an ellipsoid according to the
homogeneously collapsing ellipsoid model (HCE, solid curve), and the five
approximations considered in the text: Zel'dovich (ZA, dotted curve), 
Modified Zel'dovich (MZA, short dot), Deformation
Tensor (DTA, dash-dot), Complete Zel'dovich (CZA, short dash)
and Local Tidal (LTA, long dash). The initial axes ratios are
$1:1.25:1.5$ and the density contrast linearly extrapolated to
$a=1$ is $\delta _{0}=1$. The ZA overestimates the collapse time, whereas all other approximations are close to the HCE. \label{fig:EllipsPlot}}
\end{figure}

In Figure \ref{fig:EllipsPlot} we compare the time evolution of
the axes $R_i = aY_i$ of an ellipsoid in the five approximations
discussed in this paper. Here, the initial values $X_i$ of the
axes were arbitrarily chosen to be $1:1.25:1.5$ with $\delta
_{0}=1$. The general conclusion does not depend substantially on
the choice of these values, as will be seen in the next section.
We see that the results of these approximations, except for the
ZA, are very close to the one given by the homogeneous ellipsoid
model. The ZA overestimates the collapse time, showing that a
simple extrapolation of the linear trajectories underestimates the
nonlinear effects. The common feature we observe in the local
approximations is that the collapse occurs a little bit earlier in
the directions of the two initially larger axes than the HCE case,
whereas the collapse in the direction of the shortest axis is
slightly delayed compared to the HCE. In other words, in the local
approximations, the tidal forces are reduced compared with the HCE
model.

Concerning the collapse time $a_{c}$, all these approximations give very
similar results as shown in Table \ref{tab:colltime}. The differences are
less than $%
5\%$. The MZA gives the closest value to that of the HCE. Considering,
however, that the HCE itself neglects the effect of the interaction of the
background with the ellipsoid, this will not necessarily indicate that the
MZA has the better performance among the other local approximations. In
fact, for larger shear, the MZA deviates from the others as will be
seen in the next section.

\begin{deluxetable}{cr}
\tablecaption{\label{tab:colltime}Collapse time for a
homogeneous ellipsoid}
 \tablewidth{5cm} \tablehead{
\colhead{Approximation} & \colhead{$a_{c}$} } \startdata HCE &
$1.569$ \\ MZA & $1.578$
\\ CZA & $1.582$ \\ LTA & $1.612$ \\
DTA & $1.633$
\enddata
\end{deluxetable}

\subsection{Generic Initial Conditions\label{General}}

Following Bertschinger \& Jain (1994) we will parametrize the initial
conditions in the following way

\begin{equation}
\lambda _{i}^{0}=\frac{2}{3}\varepsilon _{0}Q_{i}\left( \alpha _{0}\right) +%
\frac{1}{3}\delta _{0},  \label{Parametrization}
\end{equation}
where $Q_{i}\left( \alpha \right) $ are the diagonal terms of the
trace\-less quadrupole matrix
\begin{equation}
Q_{ij}\left( \alpha \right) =\mathrm{diag}\left[ \cos
\left( \frac{\alpha +2\pi }{3}\right) ,\cos \left( \frac{\alpha -2\pi }{3}%
\right) ,\cos \left( \frac{\alpha }{3}\right) \right] .
\end{equation}
It is easy to show
that $\varepsilon _{0}$ is related to the magnitude of the shear and tide, $%
Q_{i}\left( \alpha _{0}\right) $ gives ratios of the eigenvalues of $E_{ij}$
and $\sigma _{ij}$ (note that in the linear regime $E_{ij}\propto \sigma
_{ij}$), and $\delta _{0}$ the density contrast. The parameter $\varepsilon
_{0}$ varies from $0$ to $\infty $, $\alpha _{0}$ from $0$ to $\pi$, and 
$\delta _{0}$ can go from $-\infty $ to $+\infty$. 
However, it is sufficient to study the dynamics for $\delta _{0}=+1$ and $%
\delta _{0}=-1$, as we shall see below.

The initial perturbation $\delta _{0}$ is defined as the ratio between $%
\delta $ and the growth factor $D$ in the linear regime:
\begin{equation}
\delta _{0}=\lim_{a\rightarrow 0}\frac{\delta }{D}.
\end{equation}
In an Einstein-de Sitter universe we have $D=a$. Thus choosing different
values of $\delta $ is equivalent to rescaling $a$. This is so for all the
kinematical parameters. Therefore, the equations of motion in the local
approximations are invariant under the following scaling
\begin{equation}
\delta _{0}\rightarrow c\delta _{0},\quad \varepsilon _{0}\rightarrow
c\varepsilon _{0},\quad \mathrm{and\quad }a\rightarrow c^{-1}a.
\label{reescaling}
\end{equation}
Due to this invariance we can express the collapse time $a_{c}$ as
(Audit, Teyssier, \& Alimi 1997):
\begin{equation}
a_{c}\left( \delta _{0},\varepsilon _{0},\alpha _{0}\right)
=
\begin{cases}
  \left| \delta _{0}^{-1}\right| a_{c}^{+}\left(
\varepsilon _{0}/\delta _{0},\alpha _{0}\right) , & \text{if
$\delta_0
>0$},\\
\left| \delta _{0}^{-1}\right| a_{c}^{-}\left( \varepsilon
_{0}/\left| \delta _{0}\right| ,\alpha _{0}\right) ,& \text{if
$\delta_0 <0$},
\end{cases}
\label{a2v}
\end{equation}
where $a_{c}^{\pm }\left( \varepsilon _{0}/\delta _{0},\alpha _{0}\right)
=a_{c}\left( \pm 1,\varepsilon _{0}/\left| \delta _{0}\right| ,\alpha
_{0}\right) $. Hence we just need to compute the two functions $a_{c}^{+}$
and $a_{c}^{-}$ which depend on $\varepsilon _{0}/\delta _{0}$ and $\alpha
_{0}$ only.

\begin{figure}
\figurenum{2}
\plottwo{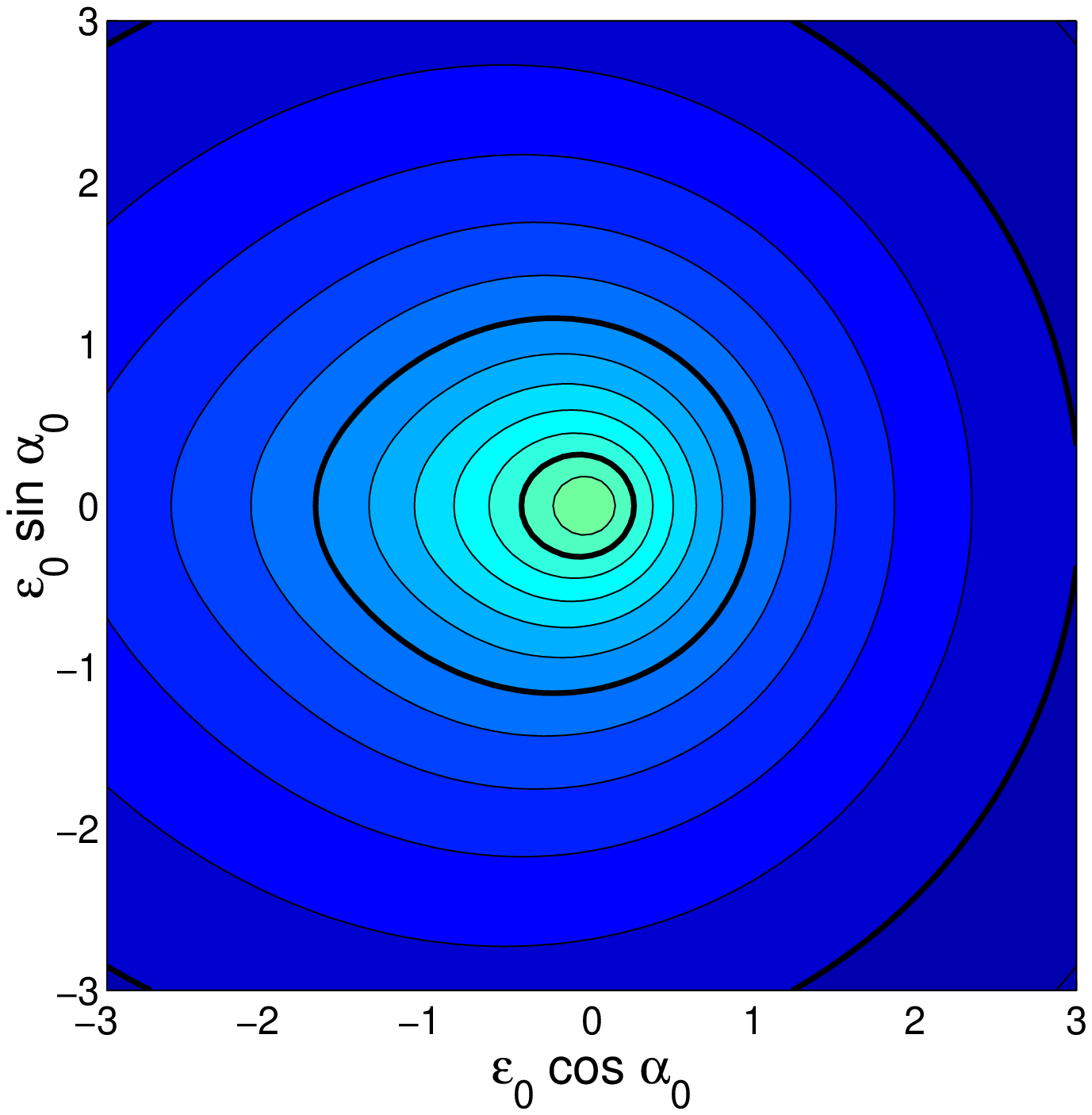}{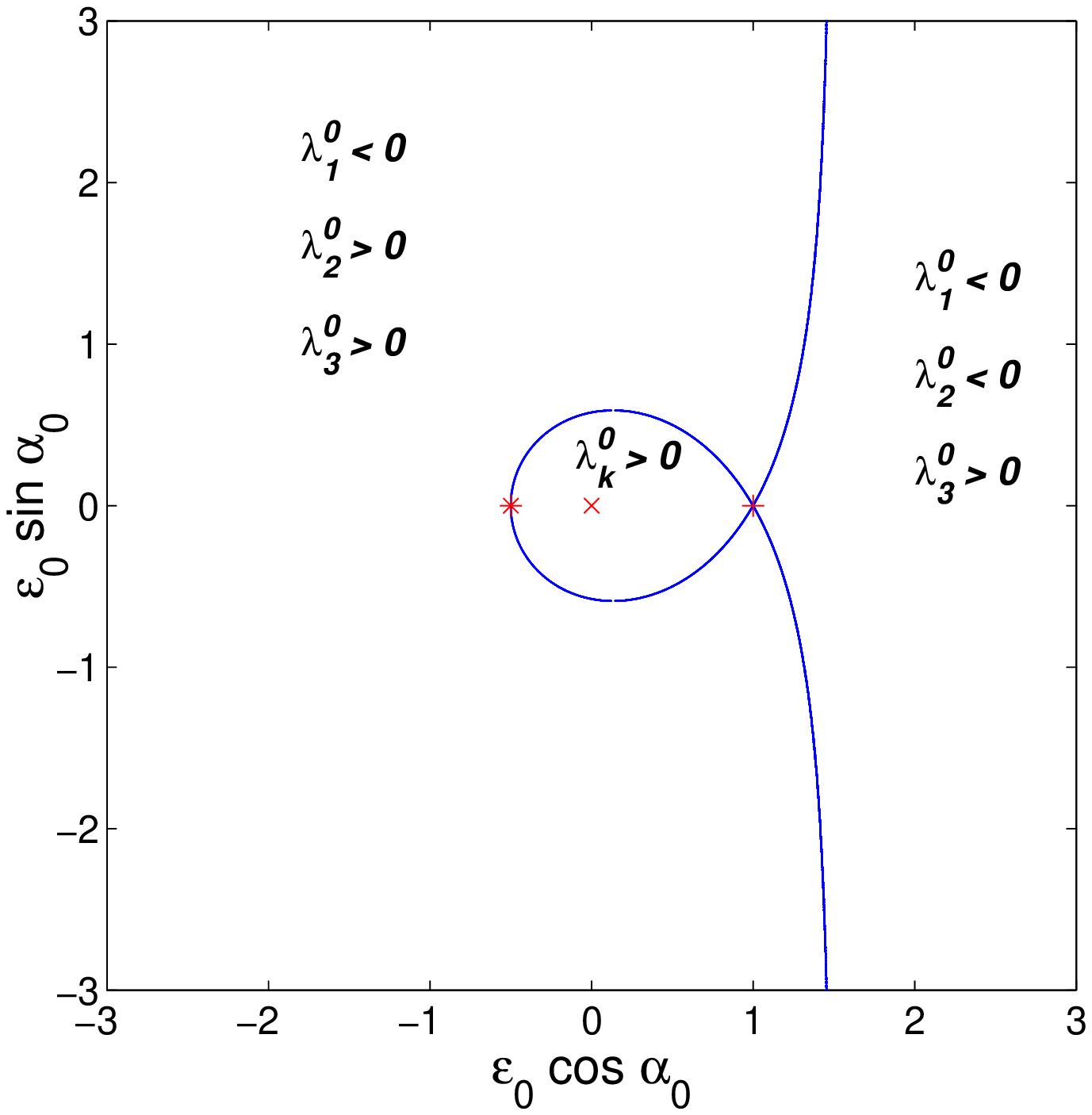}
\caption{(c) The same as (a) except that the DTA is used. The innermost contour is $a_c=1.6$. (d) The same as (a) except that the CZA is used. The innermost contour is $a_c=1.7$.
(e) The same as (a) except that the LTA is used. The innermost contour is $a_c=1.6$. (f) The signs of $\lambda^0_i$ in each region of the parameter space of initial conditions, with $\delta_0=+1$. The inner region corresponds to the values of $\lambda^0_i$ that can be spanned by a homogeneous ellipsoid with any axes ratios. Spherical, planar, and cylindrically symmetric perturbations are marked with a cross, a plus sign, and a star respectively. The upper part of these graphs, which correspond to $0 \leq \alpha_0 \leq \pi$, have $\lambda^0_3 > \lambda^0_2 >  \lambda^0_1$. The lower one covers the same values of $\lambda^0_i$, through a permutation of the indices. \label{fig:DeltaP}}
\end{figure}

\begin{figure}
\plottwo{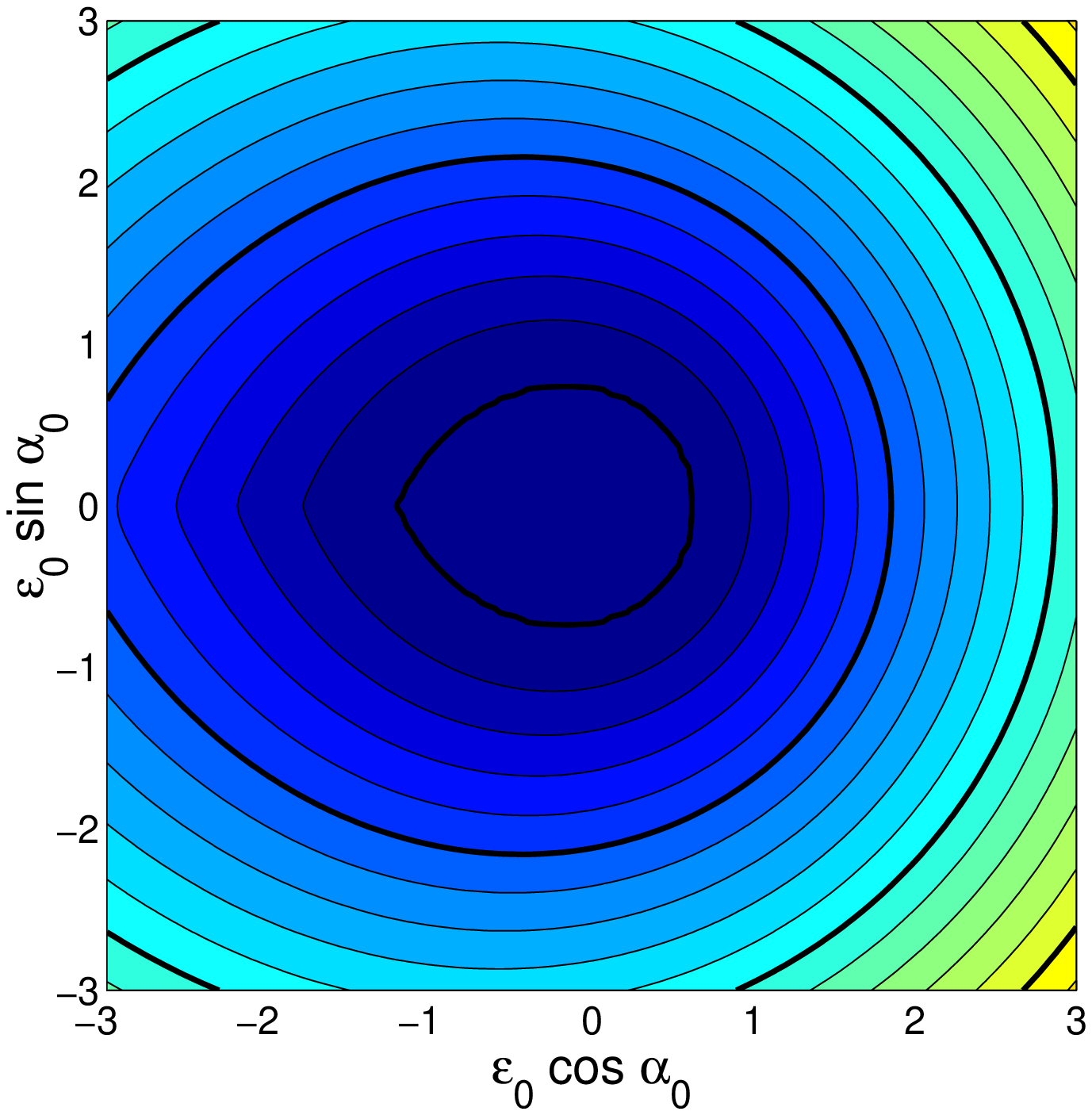}{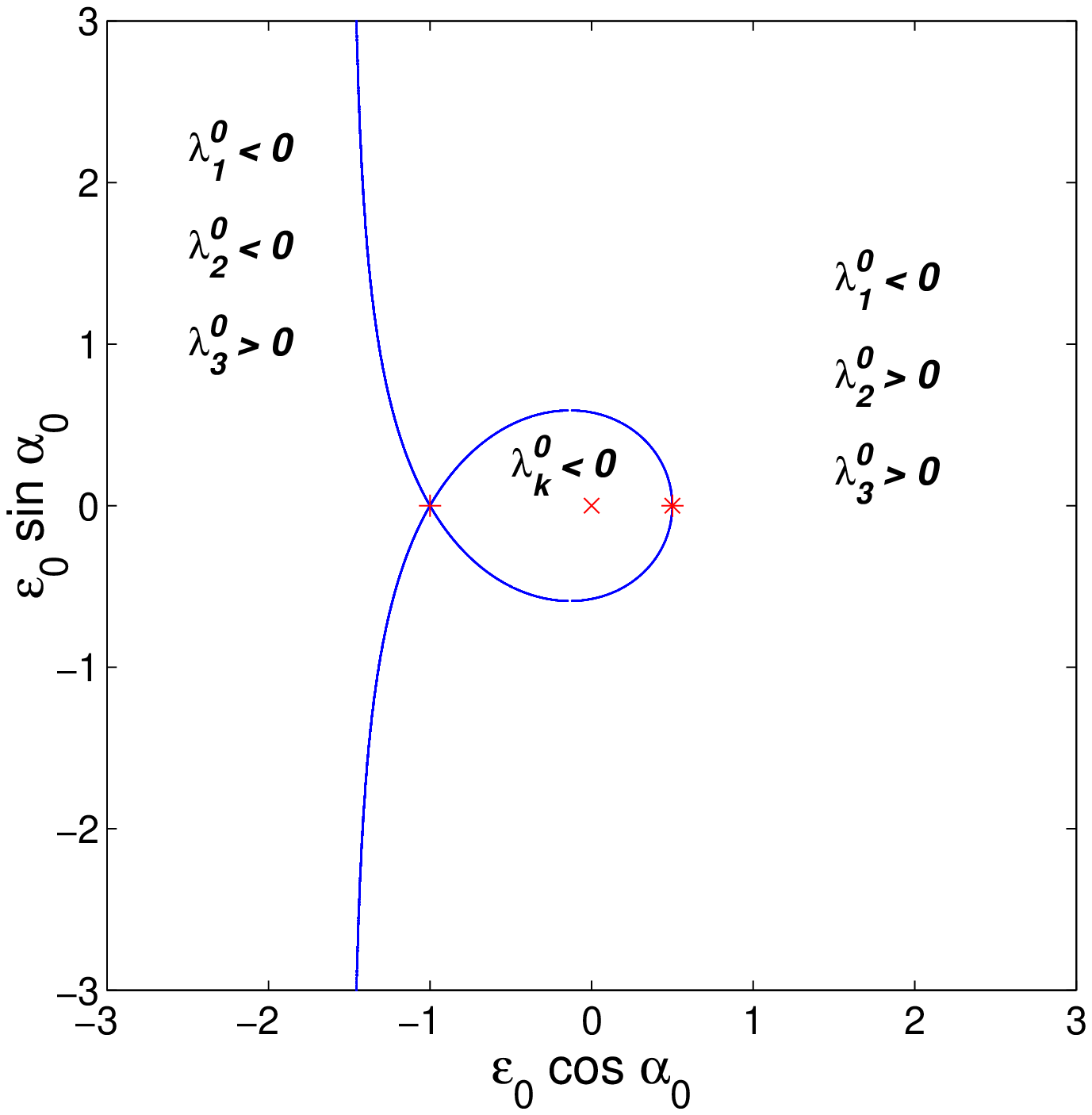}
\caption{(a) The collapse time as a function of the initial conditions for underdense perturbations with $\delta_0=-1$. The contours of constant collapse
time, expressed by $a_c^{-1}$, are displayed for the ZA. The light
(heavy) contours are spaced by 0.1 (0.5) in $a_c^{-1}$, with the innermost contour being $a_c^{-1}=0$. Initial perturbations in the central region do not collapse. (b) The same as (a) except that the DTA is used. (c) The same as (a) except that the LTA is used. (d) The signs of $\lambda^0_i$ corresponding to each region of the parameter space of initial conditions, with $\delta_0=-1$. Spherical, planar, and cylindrically symmetric underdense perturbations are marked with a cross, a plus sign, and a star respectively. The upper part of these graphs, which correspond to $0 \leq \alpha_0 \leq \pi$, have $\lambda^0_3 > \lambda^0_2 >  \lambda^0_1$. The lower one covers the same values of $\lambda^0_i$, through a permutation of the indices. \label{fig:DeltaN}}
\end{figure}

In Figures \ref{fig:DeltaP} and \ref{fig:DeltaN}, we plot the collapse
time $%
a_{c} $ as a function of $\varepsilon _{0}$ and $\alpha _{0}$ for
overdense and underdense perturbations, respectively. Since the MZA and
CZA do not apply for some underdense regions, we have not displayed the results of these approximations in figure \ref{fig:DeltaN}. We also show the signs of $\lambda^0_i$ corresponding to the initial conditions in these two figures.

The parameter space of initial conditions that can be spanned by
an ellipsoid with any axes ratios is equivalent to having the
three $\lambda^0_i$ positive. The region corresponding to the
homogeneous ellipsoid is limited to relatively small shear, and
all the local approximations, except for the ZA, agree
significantly well in this region. The ZA overestimates the collapse 
time for spherical configurations. We see that they still quite similar for
overdense perturbations in general. The MZA substantially deviates from the
others for high shear. In all the cases the shear accelerates the
collapse, which is a well-known nonlinear effect. Thus the first
regions to collapse are not necessarily those with higher density.
We can also see that oblate initial configurations 
(for which $\cos \alpha_0 > 0$) collapse first.
Thus planar collapse is favored by these approximations.

For negative perturbations the difference among the approximations is
enhanced. The LTA\ systematically gives slightly larger collapse times than
the DTA. The collapse time given by the ZA is the shortest among the three.
It is important to notice that in the local approximations underdense
regions may also collapse, due to the effects of the shear. 

The relevance of the shear in the nonlinear phase of gravitational clustering
is in agreement with $N$-body simulations (Katz, Quinn, \& Gelb 1993), yet it is sometimes ignored
in structure formation studies. Any model based on the spherical collapse
would miss this effect.

An interesting aspect of the local approximations is that the collapse time
has the same asymptotic behavior
\begin{equation}
a_{c}\left( \delta _{0},\varepsilon _{0},\alpha _{0}\right) \simeq \frac{C}{%
\varepsilon _{0}}  \label{asympt}
\end{equation}
for high initial shear ($\varepsilon _{0}\gg \delta _{0}$) in all
the approximations, where $C$ is a (slowly varying) function of
$\alpha _{0}$ only (see Appendix \ref{fits}). That is, the
collapse time for large shear does not depend on $\delta _{0}$ and
is inversely proportional to the initial shear $\varepsilon _{0}$.

\subsection{The Cosmological Mass Function}

\label{MF}

The mass function $n\left( M\right) $ is defined such that $n\left( M\right)
dM$ gives the number density of collapsed dark matter clumps with masses
between $M$ and $M+dM$. These clumps are associated with proto-galactic
haloes, and with galaxy groups and clusters. Comparing theoretical mass
functions with observations provides important constraints on the
cosmological parameters (Bahcall \& Cen 1993; Girardi et al. 1998; Rahman \&
Shandarin 2000) and the spectrum of primordial perturbations (Lucchin \&
Matarrese 1988; Ribeiro, Wuensche, \& Letelier 2000). The approach of Press
\& Schechter (1974) to calculate the mass function, hereafter PS, has been
extended to nonspherical collapse and applied to some local approximations
(Monaco 1995 for the ZA, and Audit et al. 1997 for the DTA). Here, we extend
such analysis to the LTA and compare them.

Let $F\left( M;a_{0}\right) $\ be the fraction of collapsed objects at $%
a_{0} $ with mass higher than $M$; then the mass function is given by
\begin{equation}
n\left( M\right) =-\frac{\bar{\rho}}{M}\frac{dF}{dM}.  \label{defn}
\end{equation}
The fraction $F$ may be calculated as an integral over all the possible
initial conditions weighted by their probabilities:

\begin{equation}
F=\frac{1}{F_{0}}\int_{0}^{\pi }\int_{0}^{\infty }\int_{-\infty }^{+\infty
}s\left( a_{0};\delta _{0},\varepsilon _{0},\alpha _{0}\right) P_{M}\left(
\delta _{0},\varepsilon _{0},\alpha _{0}\right) d\delta _{0}d\varepsilon
_{0}d\alpha _{0}.  \label{ncollapsed}
\end{equation}
The function $s$ is equal to one if an element with parameters $\delta
_{0},\varepsilon _{0},\alpha _{0}$ has already collapsed at $a_{0}$, and is
zero otherwise; $F_{0}$ is a normalization factor. The collapse time $a_{c}$
of a fluid element with initial perturbations parametrized by $\delta
_{0},\varepsilon _{0},\alpha _{0}$ can be computed in the local
approximations. As mentioned in subsection \ref{features}, the collapse is
characterized by the divergence of the density, which is equivalent to the
first axis collapse. Beyond this point the Lagrangian formalism breaks down.
Some authors (Audit et al. 1997; Lee \& Shandarin 1998; Sheth, Mo \& Tormen
1999) have suggested other alternatives for the definition of collapse in
the calculation of the mass function. Here we will prefer to keep the
simplest assumption of first axis collapse, since it does not introduce any
free parameter.

What we need now is the probability distribution function $P_{M}\left(
\delta _{0},\varepsilon _{0},\alpha _{0}\right) $ for the initial
conditions. Assuming Gaussian initial fluctuations, Doroshkevich (1970)
derived the joint probability for the three eigenvalues of the deformation
tensor $\lambda _{1}^{0},\lambda _{2}^{0}$ and $\lambda _{3}^{0}$. Using
this result, $P_{M}\left( \delta _{0},\varepsilon _{0},\alpha _{0}\right) $
is given by the product of three independent probabilities for each
parameter $\delta _{0}$, $\varepsilon _{0}$ and $\alpha _{0}$:

\begin{eqnarray}
P_{\nu }\left( \delta _{0}\right) &=&\frac{1}{\sqrt{2\pi \Delta ^{2}}}\exp %
\left[ -\frac{1}{2}\left( \frac{\delta _{0}}{\Delta }\right) ^{2}\right] ,
\label{Pnu} \\
P_{\chi }\left( \varepsilon _{0}\right) &=&\frac{50}{3}\sqrt{\frac{5}{2\pi
\Delta ^{2}}}\left( \frac{\varepsilon _{0}}{\Delta }\right) ^{4}\exp \left[ -%
\frac{5}{2}\left( \frac{\varepsilon _{0}}{\Delta }\right) ^{2}\right] ,
\label{Pchi} \\
P_{\alpha }\left( \alpha _{0}\right) &=&\sin \left( \frac{\alpha _{0}}{3}%
\right) \left[ \frac{3}{2}-2\sin ^{2}\left( \frac{\alpha _{0}}{3}\right) %
\right] .  \label{Pa}
\end{eqnarray}
The variance $\Delta $ is related to the mass $M$ and the power spectrum 
of the primordial density field $\sigma _{k}$ through
\begin{equation}
\Delta ^{2}\left(R\right) =\int_{0}^{\infty }\frac{dk}{2\pi ^{2}}%
W_{k}^{2}\left( R\right) k^{2}\sigma _{k}^{2} ,
\end{equation}
where $M=(4\pi /3)f_{W}R^{3}\bar{\rho}$ and $W_k\left( R\right)$ is the
Fourier transform of a filter with width $R$ in physical space. The factor $%
f_{W}$ depends on the shape of the filter function, for a top-hat filter we
have $f_{W}=1$, whereas for a sharp$-k$ filter $f_{W}=9\pi /2$. The mass 
function can now be written in the form
\begin{equation}
n\left( M\right) =-\frac{\bar{\rho}}{M}\frac{d\Delta }{dM}\Phi \left( \Delta
\right) ,  \label{golden}
\end{equation}
where $\Phi \left( \Delta \right) =dF(\Delta )/d\Delta $. The function $\Phi
\left( \Delta \right) $ contains all the influence of the dynamics and
depends neither on the particular form of the power spectrum nor of the
filter $W$; it is referred to as the universal mass function (Audit et al.
1997).

We calculate the universal mass function for the ZA, LTA and DTA but not for
the MZA and CZA since they do not apply for negative density perturbations.
In appendix \ref{calcMF} we show the detailed calculation.

\begin{figure}
\epsscale{0.8}
\plotone{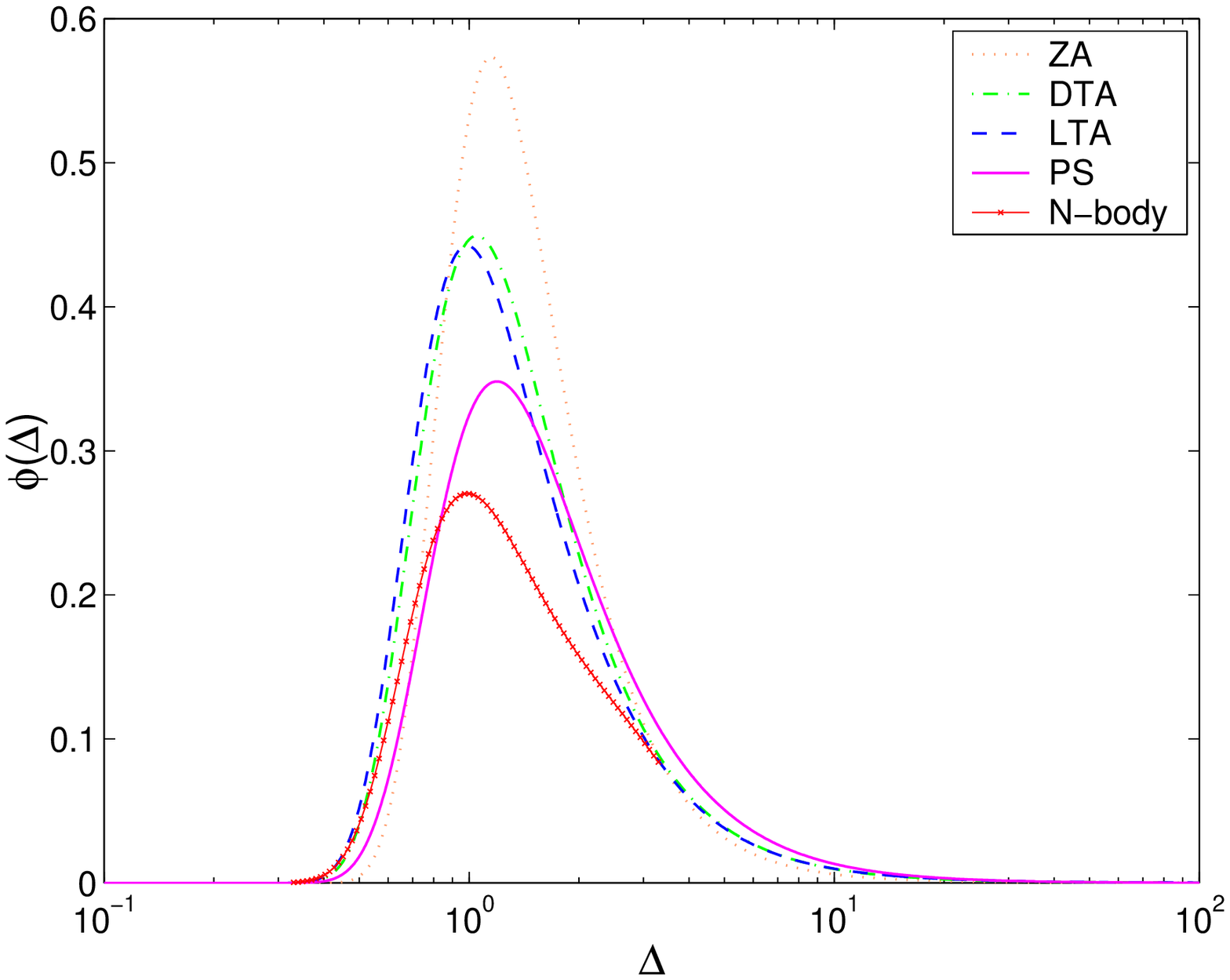}
\plotone{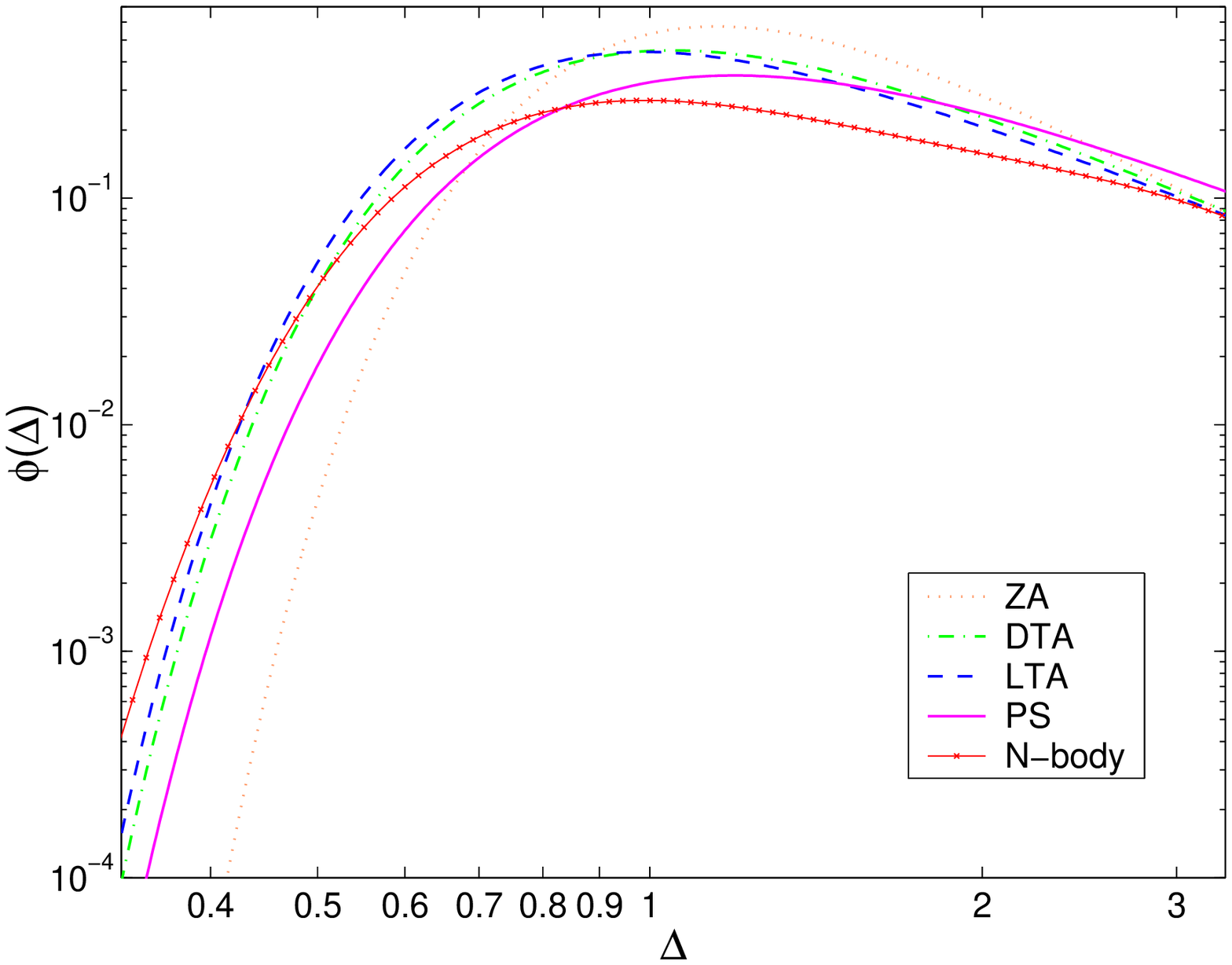}
\caption{(a) The universal mass function calculated for the ZA (dotted curve), the DTA (dash-dotted curve) and the LTA (dashed curve). For
comparison, we display in this figure the fit to
$N$-body simulations (line+cross) obtained by Jenkins et al.
(2000), together with the standard PS mass function (solid curve). (b) The same as (a) but now a logarithmic scale is used in the $y$ axis. Here the $x$ axis scale is limited to the range $0.332 \leq \Delta \leq 3.32$ covered by the $N$-body simulations. \label{fig:MFplot}}
\end{figure}

In Figure \ref{fig:MFplot}, we show the mass functions 
for those approximations. For comparison, we also display
in this figure the fit to $N$-body simulations obtained by
Jenkins et al. (2000), together with the standard PS mass
function.

We see that the results of the DTA and LTA\ are very similar.
Furthermore, in the high-mass tail ($\Delta \lesssim 0.5$), they
reproduce well the results of the $N$-body simulations. However,
we can see that these approximations overestimate the
concentration of masses near $\Delta =1$. The right-end tail of
the distribution decays more rapidly compared to the $N$-body
simulations. This tendency is still enhanced in the ZA. However,
in these approximations, the position of the maximum of the
distribution is close to that of the $N$-body simulations, giving
a better estimate than that of the PS; in particular, the LTA and DTA
give nearly the same value as the $N$-body results.

As for the normalization factor $F_{0}$, there exists an extensive
discussion on its origin (see, for example, Peacock \& Heavens 1990; Bond
et al. 1991; Jedamzik 1995; Yano, Nagashima, \& Gouda 1996). The
normalization factors for the local approximations are close to one ($1/0.92$
for the DTA, and $1/0.89$ for the LTA) whereas in the original PS derivation
the normalization factor needed is $F_{0}=1/0.5$. This is due to the fact
that, in the spherical collapse model, only overdense regions collapse.

The fact that around $\Delta = 1$ the number of objects is
overestimated in the local approximations implies, due to the
normalization of the mass function, that they should provide a
lower estimate than the $N$-body simulations for large enough
$\Delta$. There, the contribution from the low-mass objects is
dominant; in any realistic process, they may also arise from the
fragmentation of larger clusters. The criterion for the formation
of a clump from the direct collapse of an initially perturbed
region does not account for these complex processes of
fragmentation. Therefore, the discrepancy of the mass function for
high $\Delta $ might be attributed to the use of expression
(\ref{ncollapsed}) rather than to the definition (\ref{defn}).

\section{DISCUSSION}

\label{conclusion}

We have investigated local Lagrangian approximations to the nonlinear
dynamics of pressureless dark matter. We have selected the modified
Zel'dovich approximation (MZA), the deformation tensor approximation
(DTA), the complete Zel'dovich approximation (CZA), and the local tidal approximation (LTA), in addition to the original Zel'dovich approximation (ZA). These four approximations were designed to improve the ZA, and are in fact  
exact for planar, spherical and cylindrical symmetries, whereas the ZA is
only exact for the planar case. They are semi-analytic and easy to be
implemented in any application where local quantities are involved, such as
the calculation of the mass function in the PS approach.

All the local approximations discussed here, except for the ZA, provide
quite a similar evolution for an ellipsoid, reproducing the results of
the homogeneous ellipsoid model. Thus, for this kind of positive density
perturbations, these methods work fairly well. However, the MZA turns out to
deviate substantially for large values of the shear as was shown in section (\ref{General}), reflecting the fact that it does not give the correct second
order solution. Furthermore, the MZA cannot deal with initially underdense
regions that will eventually collapse. Therefore, its applicability is
rather limited when compared to the other approximations.

We note that the second order expansions of the CZA, LTA and DTA coincide
with the second order Lagrangian perturbation theory, whereas those of the
MZA and ZA do not. It is interesting to recall that the LTA and DTA have
very different origins from the CZA, but still they give the correct second
order result.

The CZA, LTA and DTA give quite analogous results for generic initial
conditions. However, at least in its original form, the CZA cannot be used
for negative values of $\lambda _{i}$. One possible solution to this problem
might be achieved through an expansion such as
\begin{equation}
r_{i}=\frac{1+\sum_{l,m,n=0}^{N}E_{l,m,n}^{p}(\lambda _{j}+\lambda
_{k})^{l}(\lambda _{j}-\lambda _{k})^{2n}\lambda _{i}^{m}}{%
1+\sum_{l,m,n=0}^{M}D_{l,m,n}^{p}(\lambda _{j}+\lambda _{k})^{l}(\lambda
_{j}-\lambda _{k})^{2n}\lambda _{i}^{m}}.
\end{equation}
The coefficients $E$ and $D$ should be appropriately chosen to
adjust the asymptotic behavior; in particular, we could use the
numerical solution for underdense cylindrical and spherical
perturbations to fit some of these coefficients, as done for the
overdense case in the CZA. Besides, to agree with the perturbative
solution (\ref{CZA}) we should have
\begin{equation}
E_{l,m,n}^{1}-D_{l,m,n}^{1}=C_{l,m,n}^{1}.
\end{equation}
For higher orders, the determination of these coefficients is
rather complicated. Further investigations on this possibility
should be pursued.

Concerning the mass function, it is found that the LTA and DTA
give an accurate result for large masses as compared to the
$N$-body simulations. The position of the peak is also in good
agreement, whereas its amplitude is overestimated by a factor 2
compared to the $N$-body results. Since the mass function is
normalized to unity, this means that the local approximations,
together with the PS formalism, underestimates the density of
low-mass clusters. However, this might be a consequence of the
criterion for the formation of a collapsed object based only on
the collapse time.

It is interesting to notice that the collapse time, as a function of $\alpha
_{0}$ and $\varepsilon _{0},$ has an approximate scaling property (see eq.
\ref{separability}), which is very precise for $\varepsilon _{0}/\delta
_{0}\gg 1$. We conclude that this may be a general feature of gravitational
collapse in local approximations, whose validity is worth checking in a more
general setting.

While there is still not a clear theoretical understanding or
support to the local approximations, they proved to be very
accurate in the situations investigated here. They reproduce some
well known features of nonspherical collapse, such as the
possibility of collapse of some initially underdense regions, and
the fact that the shear accelerates the collapse (see Sahni \&
Coles 1995). The main limitation of these approximations is that
they only provide information about the internal state of a given
mass element, but do not determine its position. Even so, their
simplicity is highly expedient for practical applications, such as the
calculation of nonlinear corrections to the microwave background
anisotropies, and the Gunn-Peterson effect. In particular, they are
suitable for obtaining statistical properties of the present
fields as a function of the primordial ones, as in the case of the
mass function. Further studies on the validity of these
approximations, based on a comparison to $N$-body simulations, are
required. Such a comparison would allow to fully test the
approximations described in this paper and more generically, the
locality hypothesis. If they still provide accurate results in
this case, the local approximations could represent good
alternatives to the computer simulations, taking much less
computational time, allowing thereby a larger scanning of initial
conditions. They could give complementary information to the
$N$-body simulations and would provide a better physical
understanding of the nonlinear dynamics of self gravitating
systems.

Most of the results of this paper may be extended to more general
backgrounds. The influence of any smooth component only alters 
the behavior of $a(t)$, and of the growing mode growth factor $D\left( t\right)$,
that will not be equal any more. It would be interesting to
study the relativistic analogue of the LTA. Another interesting
extension would be to include vorticity in the local
approximations as was done for the ZA in (Buchert 1992; Barrow \&
Saich 1993). We could use them to test the effects of a possible
primeval vorticity on large scales (Li Xin-Li 1998).

\acknowledgments

MM acknowledges the participants of the ``Pequeno Semin\'{a}rio''
at CBPF, and was partially supported by a CNPq fellowship  (contract  no 142338/97-4). 
TK would like to acknowledge the partial support of the Brazilian sponsoring 
agencies CNPq (contract  no 300962/86-0) and FAPERJ 
(contract  no E-26/150.942/99), and a PRONEX grant (contract  no 41.96.0886.00).
M. O. C. acknowledges financial support from FUJB-UFRJ.

\appendix

\section{FITTING FORMULAE FOR THE COLLAPSE TIME}

\label{fits}

In order to avoid repeated numerical integrations of differential equations
in the LTA and DTA, we have parametrized the collapse time as a function of
initial conditions. For both of these local approximations, the following
scaling property is approximately satisfied:
\begin{equation}
a_{c}^{\pm }\left( x,\alpha _{0}\right) \simeq H^{\pm }\left( f\right) ,\;%
\mathrm{with\;}f=xg^{\pm }\left( \alpha _{0}\right) ,  \label{separability}
\end{equation}
where $H^{\pm }$ and $g^{\pm }$ are functions to be fitted for each
approximation, and $x=\varepsilon _{0}/\left| \delta _{0}\right| $. This
relation becomes more accurate for increasing $x$. In any case the error of
the fit is less than a few percent.

For $g\left( \alpha \right) $ we found that a kind of truncated Fourier
series can be used to a very good approximation:
\begin{eqnarray}
g\left( \alpha \right) &=&c_{1}\cos \left( \frac{\alpha
}{2}\right) +c_{2}\cos \alpha +c_{3}\cos \left( \frac{3\alpha
}{2}\right) +c_{4}\cos \left( 2\alpha \right) + \nonumber \\
&&c_{5}\cos \left( \frac{5\alpha }{2}\right) +\left(
1-c_{1}-c_{2}-c_{3}-c_{4}-c_{5}\right) \cos \left( 3\alpha \right)
.
\end{eqnarray}
The values of the parameters for each case are shown in Table
\ref{tab:gpar}.

\begin{deluxetable}{cccccc}
\tablecaption{\label{tab:gpar}Parameters of $g(\alpha)$ fitted for the LTA and DTA}
\tablewidth{8cm}
\tablehead{
\colhead{$g$} & \colhead{$c_{1}$} & \colhead{$c_{2}$} &
\colhead{$c_{3}$} & \colhead{$c_{4}$} & \colhead{$c_{5}$}
}
\startdata
$g_{LTA}^{+}$ & 1.546 & -1.015 & 0.786 & -0.462 & 0.182 \\
$g_{LTA}^{-}$ & 1.461 & -0.767 & 0.473 & -0.231 & 0.079 \\
$g_{DTA}^{+}$ & 1.505 & -0.842 & 0.508 & -0.228 & 0.066 \\
$g_{DTA}^{-}$ & 1.497 & -0.836 & 0.513 & -0.234 & 0.068 \\
\enddata
\end{deluxetable}

For the dependence of $a_{c}$ on $x$ for fixed $\alpha _{0}$, we fitted the
function $f=H^{-1}$ rather than $H$ itself, because this is the quantity we
will need to compute the mass function. Taking into account the boundary
value and asymptotic behavior, we parametrize $f$ by
\begin{equation}
f^{+}=\frac{d_{1}z\left( 1+d_{2}z+d_{4}z^{2}\right) }{\left(
1+d_{3}z+d_{5}z^{2}\right) a_{c}},\;\mathrm{with\;}z=d_{0}-a_{c},
\end{equation}
for overdense regions ($\delta _{0}>0$), and
\begin{equation}
f^{-}=d_{0}+\frac{d_{1}a_{c}^{-1/2}\left(
1+d_{2}a_{c}^{-1}+d_{4}a_{c}^{-5/2}\right) }{%
1+d_{3}a_{c}^{-1}+d_{5}a_{c}^{-2}},
\end{equation}
for underdense regions ($\delta _{0}<0$). The parameters for the
LTA and DTA are given in Table \ref{tab:fpar}.

\begin{deluxetable}{lllllll}
\tablecaption{\label{tab:fpar}Parameters of $f$ fitted for the LTA and DTA}
\tablewidth{9.5cm}
\tablehead{
\colhead{$f$} & \colhead{$d_{0}$} & \colhead{$d_{1}$} &
\colhead{$d_{2}$} & \colhead{$d_{3}$} & \colhead{$d_{4}$} &
\colhead{$d_{5}$}
}
\startdata
$f_{LTA}^{+}$ & $1.686$ & $8.469$ & $19.88$ & $78.49$ & $17.03$
& $163.2$ \\ $f_{DTA}^{+}$ & $1.686$ & $14.14$ & $13.34$ & $87.68$
& $8.676$ & $163.8$ \\ $f_{LTA}^{-}$ & $0.591$ & $1.064$ & $0.678$
& $-1.335$ & $16.054$ & $8.613$
\\
$f_{DTA}^{-}$ & $0.495$ & $0.942$ & $0.322$ & $-2.083$ & $25.718$ &
$12.961$ \\
\enddata
\end{deluxetable}

Notice that for high shear ($x\gg 1$) we have $a_{c}\ll 1$ such that:
\begin{equation}
a_{c}^{+}\underset{x\rightarrow \infty }{\longrightarrow }\frac{%
d_{1}d_{0}\left( 1+d_{2}d_{0}+d_{4}d_{0}^{2}\right) }{\left(
1+d_{3}d_{0}+d_{5}d_{0}^{2}\right) }\frac{1}{xg^{+}\left( \alpha _{0}\right)
}.
\end{equation}
and
\begin{equation}
a_{c}^{-}\underset{x\rightarrow \infty }{\longrightarrow }\frac{d_{1}d_{4}}{%
d_{5}}\frac{1}{xg^{-}\left( \alpha _{0}\right) },
\end{equation}

In the Zel'dovich approximation we have an analytical expression for $a_{c}$.
As the collapse occurs when the greatest $\lambda _{i}$ reaches the
value $1$, using the parametrization (\ref{Parametrization}) we get:
\begin{equation}
a_{c}^{ZA}=\frac{3}{\delta _{0}+2\varepsilon _{0}\cos \left( \alpha
_{0}/3\right) }.
\end{equation}
In this case we can clearly see the features of $a_{c}$:
\begin{eqnarray}
a_{c}^{ZA}\left( \delta _{0},\varepsilon _{0},\alpha _{0}\right) &=&\frac{1}{%
\left| \delta _{0}\right| }\frac{3}{\pm 1+2\left( \varepsilon
_{0}/\left| \delta _{0}\right| \right) \cos \left( \alpha
_{0}/3\right) } \nonumber \\ &=&\frac{1}{\left| \delta _{0}\right|
}a_{c}^{ZA}\left( \pm 1,\varepsilon _{0}/\left| \delta _{0}\right|
,\alpha _{0}\right) .
\end{eqnarray}
Note that the property (\ref{separability}) is satisfied exactly for the ZA.

\section{CALCULATION OF THE MASS FUNCTION}

\label{calcMF}

With the integral (\ref{ncollapsed}) we may write the universal mass
function $\Phi \left( \Delta \right) =dF/d\Delta $ in the form
\begin{equation}
\Phi \left( \Delta \right) =\frac{1}{F_{0}}\frac{d}{d\Delta }\int_{0}^{\pi
}\int_{0}^{\infty }\int_{-\infty }^{+\infty }s\left( \nu \Delta ,\chi \Delta
,\alpha _{0}\right) P\left( \nu ,\chi ,\alpha _{0}\right) d\nu d\chi d\alpha
_{0},  \label{fiDelnu}
\end{equation}
where $\nu =\delta _{0}/\Delta $ and $\chi =\varepsilon _{0}/\Delta $. With
these new variables the dependence on $\Delta $ will be present only in 
the function $s$, which may be written as
\begin{equation}
s=\Theta \left( 1-a_{c}\left( \delta _{0},\varepsilon _{0},\alpha
_{0}\right) \right) .
\end{equation}
Note that in the case of the Press \& Schechter original approach this
function is given by: $s=\Theta \left( \delta _{0}-\delta _{c}\right) ,$
where $\delta _{c}=1.686$ is the value at which a spherical perturbation
collapses at $a=1$.

To calculate $ds/d\Delta $ one uses the relation (\ref{a2v}), obtaining:
\begin{equation}
\frac{ds}{d\Delta }=\delta _{D}\left( 1-a_{c}\left( \nu \Delta ,\chi \Delta
,\alpha _{0}\right) \right) \frac{a_{c}^{\pm }\left( \chi /\nu ,\alpha
_{0}\right) }{\left| \nu \right| \Delta ^{2}},
\end{equation}
where $\delta _{D}$ is the Dirac delta function. Therefore we may
eliminate one of the integrals in expression (\ref{fiDelnu}), with
the mass function being calculated over the surface $a_{c}\left(
\nu \Delta ,\chi \Delta ,\alpha _{0}\right) =\left| \nu \Delta
\right| ^{-1}a_{c}^{\pm }\left( \chi /\nu ,\alpha _{0}\right)$
$=1$. For this sake we need to write one of the three variables in
terms of the others on this surface.

Let us assume the we have $\chi $ as a functions of $\nu \Delta $ and $%
\alpha $: $\chi =\chi _{a}\left( \nu \Delta ,\alpha \right) $, where the
subscript $a$ indicates that $\chi $ is calculated over the surface $a_{c}=1$%
. The integral in $\chi $ in equation (\ref{fiDelnu}) is thus eliminated
using the relation
\begin{equation}
\delta _{D}\left( 1-a_{c}\left( \nu \Delta ,\chi \Delta ,\alpha _{0}\right)
\right) =\delta _{D}\left( \chi -\chi _{a}\right) \left| \frac{\partial a_{c}%
}{\partial \chi }\right| ^{-1}.
\end{equation}
The mass function (\ref{fiDelnu}) is now given by:
\begin{equation}
\Phi \left( \Delta \right) =\frac{1}{F_{0}}\int_{0}^{\pi }\int_{-\infty
}^{+\infty }\frac{1}{\Delta}\left| \frac{\partial a_{c}}{\partial
\chi }\right| ^{-1}P_{\nu }\left( \nu \right) P_{\chi }\left( \chi
_{a}\right) P_{\alpha }\left( \alpha _{0}\right) d\nu d\alpha _{0},
\label{fiDelnu2}
\end{equation}  
where $\partial a_{c}/\partial \chi $ is evaluated in $\chi _{a}\left( \nu
\Delta ,\alpha \right) .$ We can simplify this expression further if the
collapse time $a_{c}\left( x,\alpha \right) $ is only a function of the
product $xg\left( \alpha \right) ,$ which we have seen is an excellent
approximation for the local approximations studied here (see appendix \ref{fits}).
Using the property (\ref{separability}) we have
\begin{equation}
\chi _{a}=\frac{\nu }{g\left( \alpha _{0}\right) }H^{-1}\left( \left| \nu
\Delta \right| \right) ,
\end{equation}
and
\begin{equation}
\left| \frac{\partial a_{c}}{\partial \chi }\right| =\left| \frac{1}{\nu
\Delta }\frac{\partial H}{\partial f}\frac{g\left( \alpha _{0}\right) }{\nu }%
\right| ,
\end{equation}
where the superscript $+$ is implied for positive $\nu $, and the $-$ for
negative $\nu $. Replacing these results in expression (\ref{fiDelnu2}) we
get finally:
\begin{equation}
\Phi \left( \Delta \right) =\frac{1}{F_{0}}\int_{-\infty }^{+\infty
}\int_{0}^{\pi }\nu ^{2}\left| \left( \frac{\partial H}{\partial f}\right)
^{-1}\frac{1}{g\left( \alpha _{0}\right) }\right| P_{\nu }\left( \nu \right)
P_{\chi }\left( \chi _{a}\right) P_{\alpha }\left( \alpha _{0}\right)
d\alpha _{0}d\nu ,
\end{equation}
where $\chi _{a}=\left| \nu \right| f_{a}/g\left( \alpha _{0}\right) $, with
$f_{a}\left( \nu \Delta \right) =H^{-1}\left( \left| \nu \Delta \right|
\right) $ and $\partial H/\partial f$, is calculated in $f_{a}$. This is why
we have chosen to fit the function $H^{-1}$, instead of its inverse.

As $P\left( \nu \right) $ and $\partial H/\partial f$ are independent of $%
\alpha _{0},$ we integrate first in this variable:

\begin{equation}
I_{1}\left( \nu ,\Delta \right) :=\int_{0}^{\pi }\frac{1}{g^{5}\left( \alpha
_{0}\right) }\exp \left[ -\frac{5}{2}\left( \frac{\nu f}{g\left( \alpha
_{0}\right) }\right) ^{2}\right] P_{\alpha }\left( \alpha _{0}\right)
d\alpha _{0}.  \label{I1}
\end{equation}
The universal mass function will now be given by
\begin{equation}
\Phi \left( \Delta \right) =\frac{1}{F_{0}}N\int_{-\infty }^{+\infty
}I_{1}\left( \nu ,\Delta \right) \nu ^{6}f_{a}^{4}\left| \left( \left. \frac{%
\partial H}{\partial f}\right| _{f_{a}}\right) ^{-1}\right| \exp \left( -%
\frac{\nu ^{2}}{2}\right) d\nu .  \label{I2}
\end{equation}
where $N=50\sqrt{5}/6\pi $ is the product of the normalizations
for $P_{\nu } $ and $P_{\chi }$ (equations (\ref{Pnu}) and
(\ref{Pchi})). Note that the above integral is limited for
positive values of $\nu $, as $H^{-1}\left( \left| \nu \Delta
\right| \right) =0$ for $\nu \Delta >f_{0}$ ($f_{0}=1.686$ for the
DTA and LTA, and $f_{0}=3$, for the ZA)$.$ Although our fitting
formulas (sec. \ref{fits}) become less accurate for $\nu \rightarrow
-\infty $ and $\nu \rightarrow f_{0}/\Delta ,$ the mass function
is not affected, since the integrand in (\ref{I2}) goes to zero in
these regions. Here it is clear that the underdense regions do
contribute to the mass function, as pointed out by Audit et al.
(1997).

As an example, let us consider the Zel'dovich approximation. In this case we
have
\begin{equation}
H_{\pm }\left( f\right) =\frac{3}{\pm 1+2f},\;\mathrm{and\;}g\left( \alpha
\right) =\cos \left( \frac{\alpha }{3}\right) .
\end{equation}
Using the following transformation of variables

\begin{equation}
x=\sin \left( \frac{\alpha }{3}\right) \rightarrow P_{\alpha }d\alpha =x%
\left[ \frac{3}{2}-2x^{2}\right] 3\frac{dx}{\sqrt{1-x^{2}}},
\end{equation}
we find an analytical expression for the integral (\ref{I1}):
\begin{equation}
I_{1}:=\frac{1}{\left( \nu f\right) ^{2}}\exp \left[ -\frac{5}{2}\left( \nu
f\right) ^{2}\right] \left[ \frac{3}{25}\frac{1}{\left( \nu f\right) ^{2}}%
\left( \exp \left[ -\frac{15}{2}\left( \nu f\right) ^{2}\right] -1\right) +%
\frac{9}{10}\right] ,
\end{equation}
where $f^{\pm }=\left( 3/\left( \left| \nu \right| \Delta \right) \mp
1\right) /2$. The mass function will be given by:
\begin{eqnarray}
\Phi \left( \Delta \right)&  =  &\frac{1}{F_{0}}\frac{15\sqrt{5}}{8\pi }%
\int_{-\infty }^{+\infty }\frac{1}{\Delta ^{4}}\exp \left( -\frac{\nu ^{2}}{2%
}\right) \exp \left[ -\frac{5}{8}\frac{\left( 3-\nu \Delta \right) ^{2}}{%
\Delta ^{2}}\right] \times \nonumber \\ & & \left[
\frac{12}{5}\Delta ^{2}\left( \exp \left[
-\frac{15}{8}\frac{\left( 3-\nu \Delta \right) ^{2}}{\Delta
^{2}}\right] -1\right) +\frac{9}{2}\left( 3-\nu \Delta \right)
^{2}\right] d\nu .
\end{eqnarray}

\clearpage

\begin{figure}
\plottwo{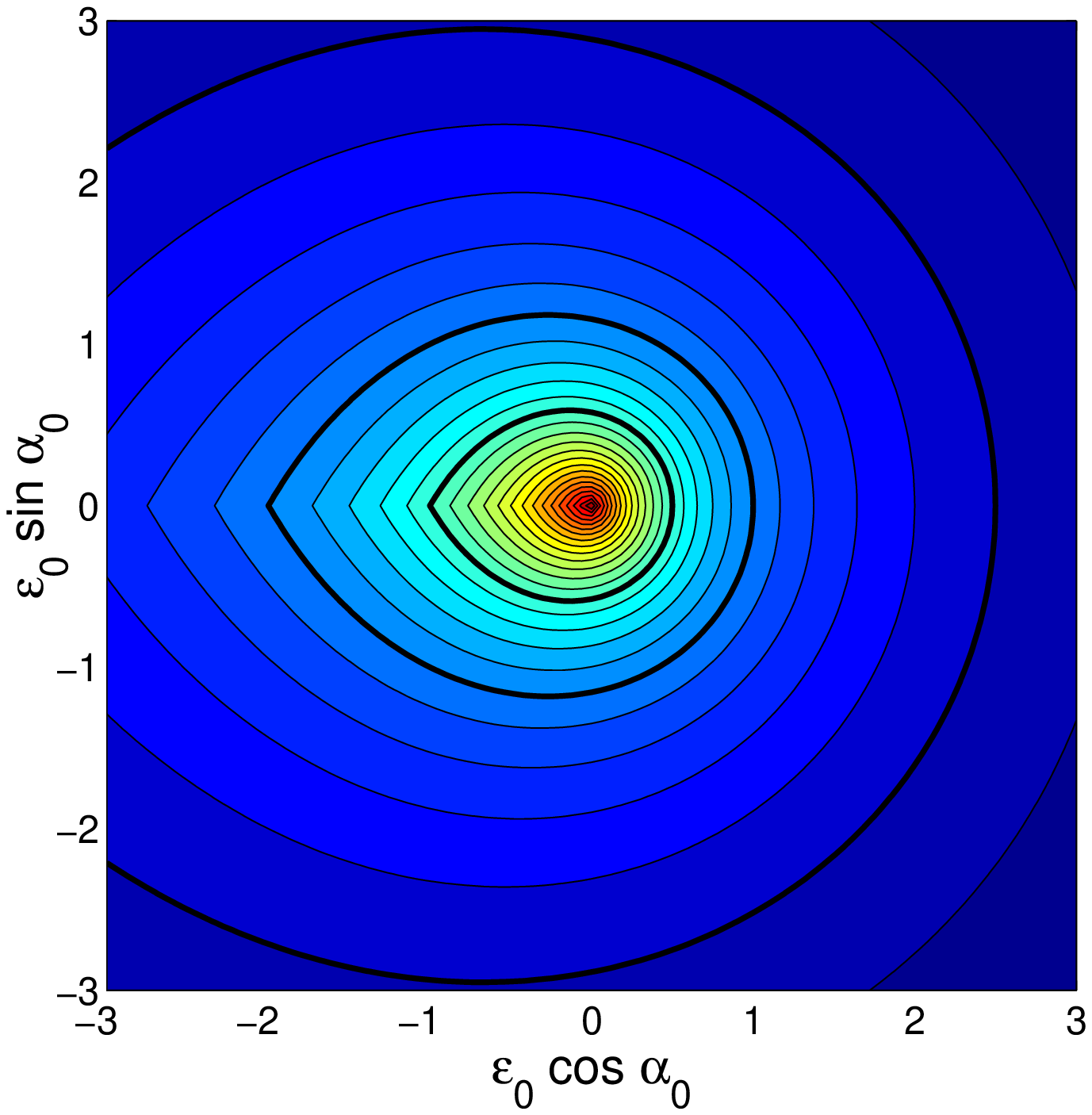}{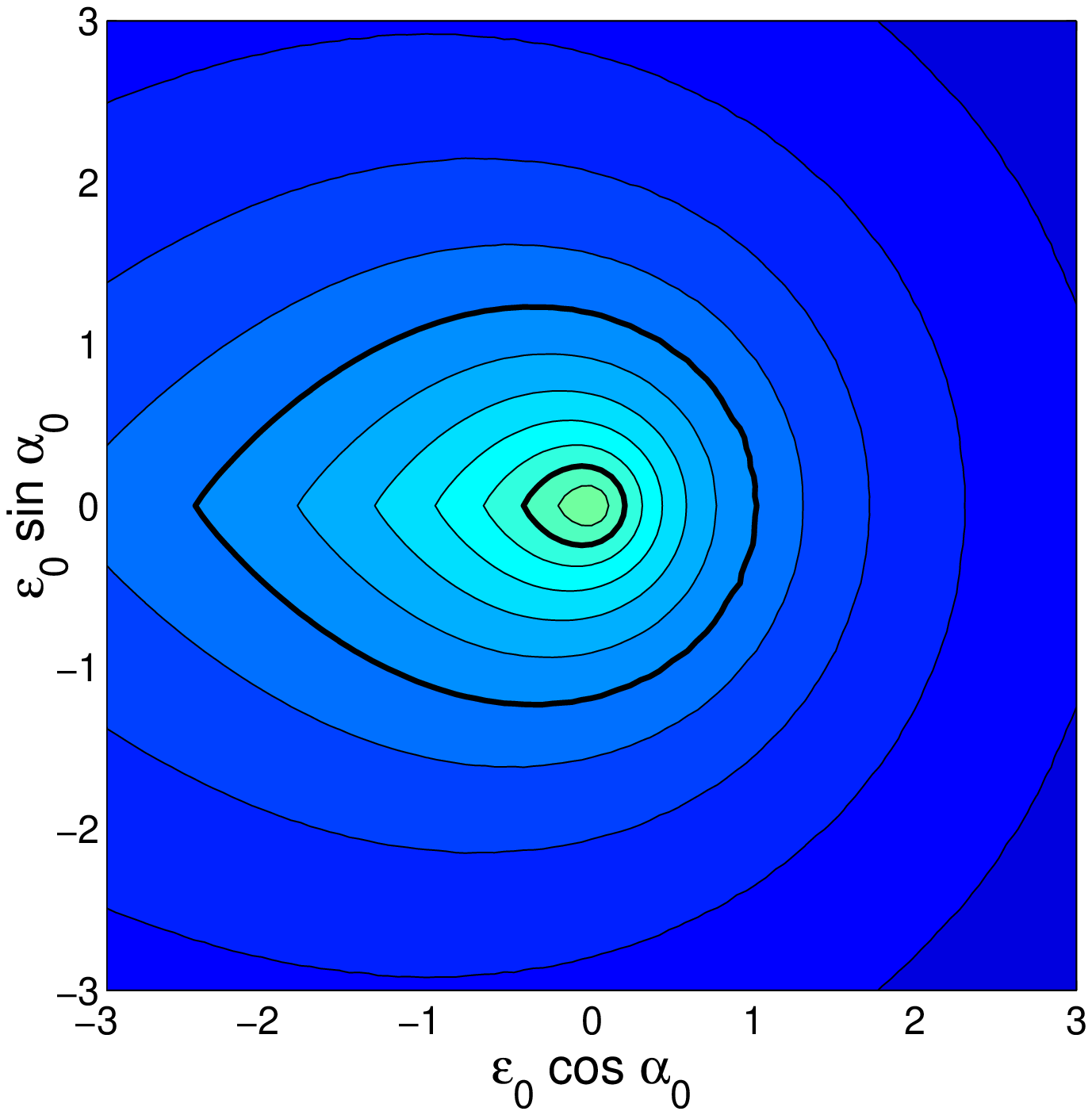}
\caption{(a) The collapse time as a function of the initial conditions for overdense perturbations with $\delta_0=+1$. The contours of constant collapse
time, expressed by the scale factor $a_c$, are displayed for the ZA. The light
(heavy) contours are spaced by 0.1 (0.5) in $a_c$, with the outermost contour
being $a_c=0.4$ and the central velue $a_c=3$. (b) The same as (a) except that
the MZA is used. The innermost contour is $a_c=1.6$.}
\end{figure}

\clearpage

\begin{figure}
\plottwo{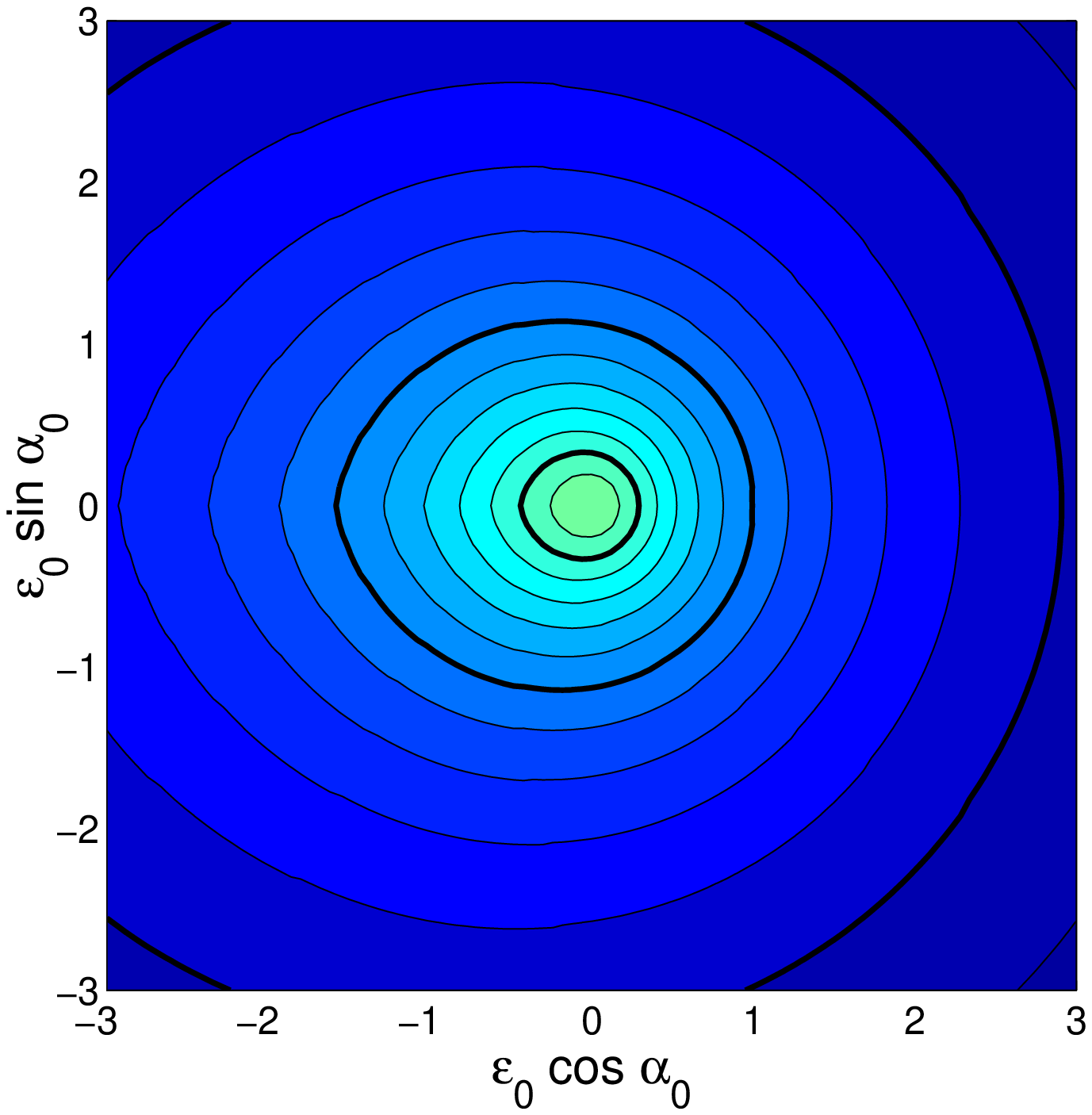}{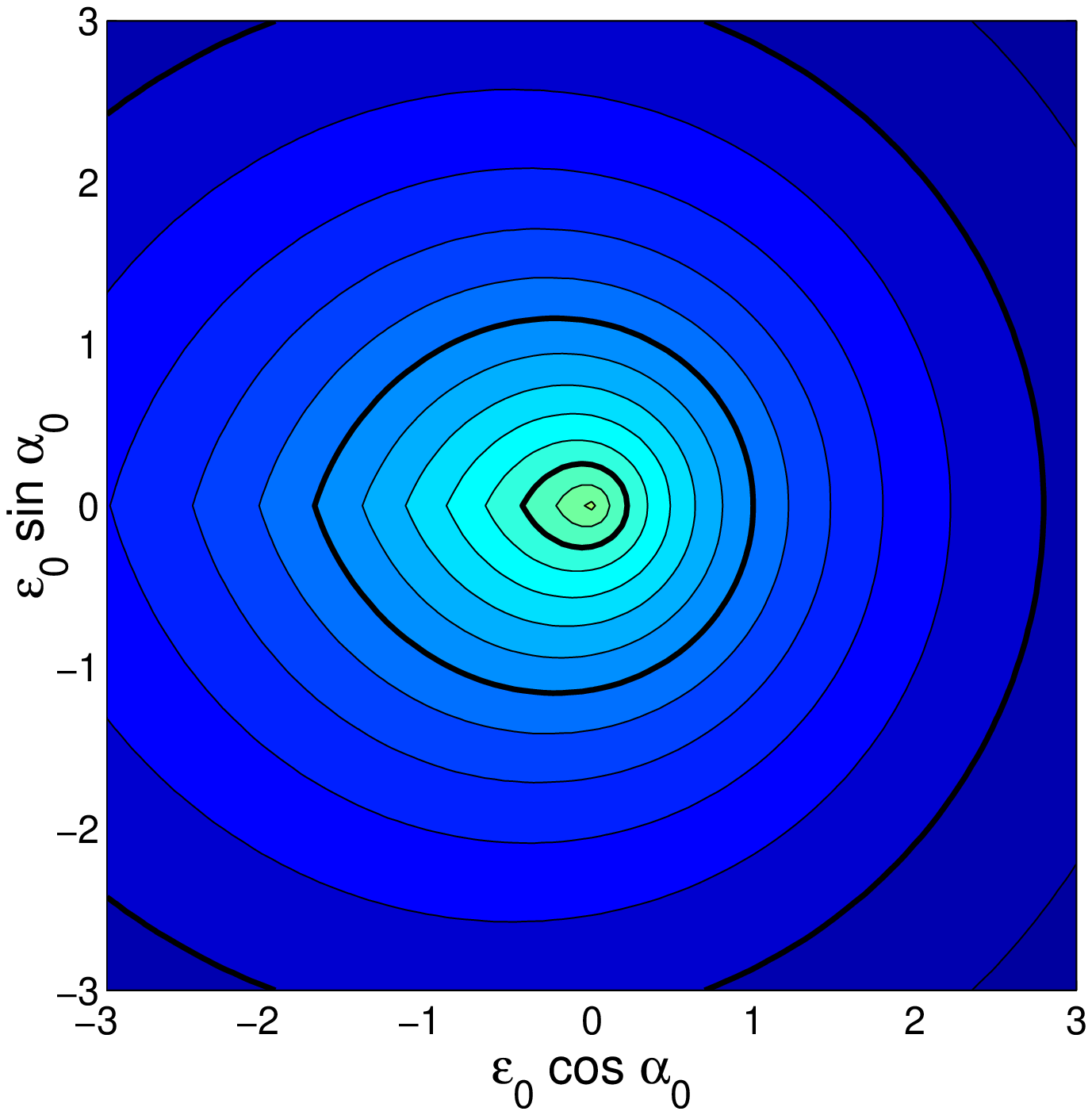}
\end{figure}

\clearpage

\begin{figure}
\plottwo{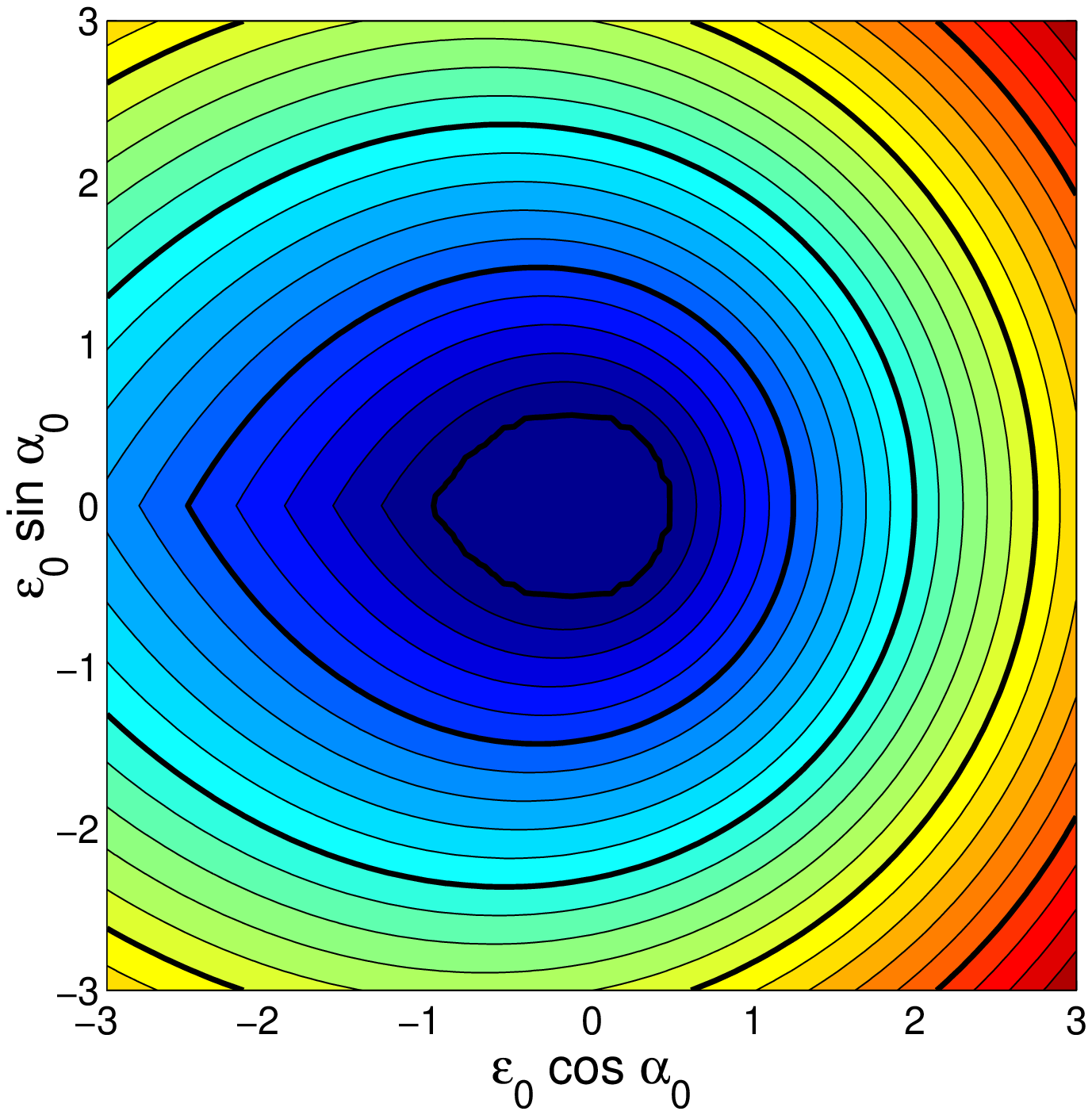}{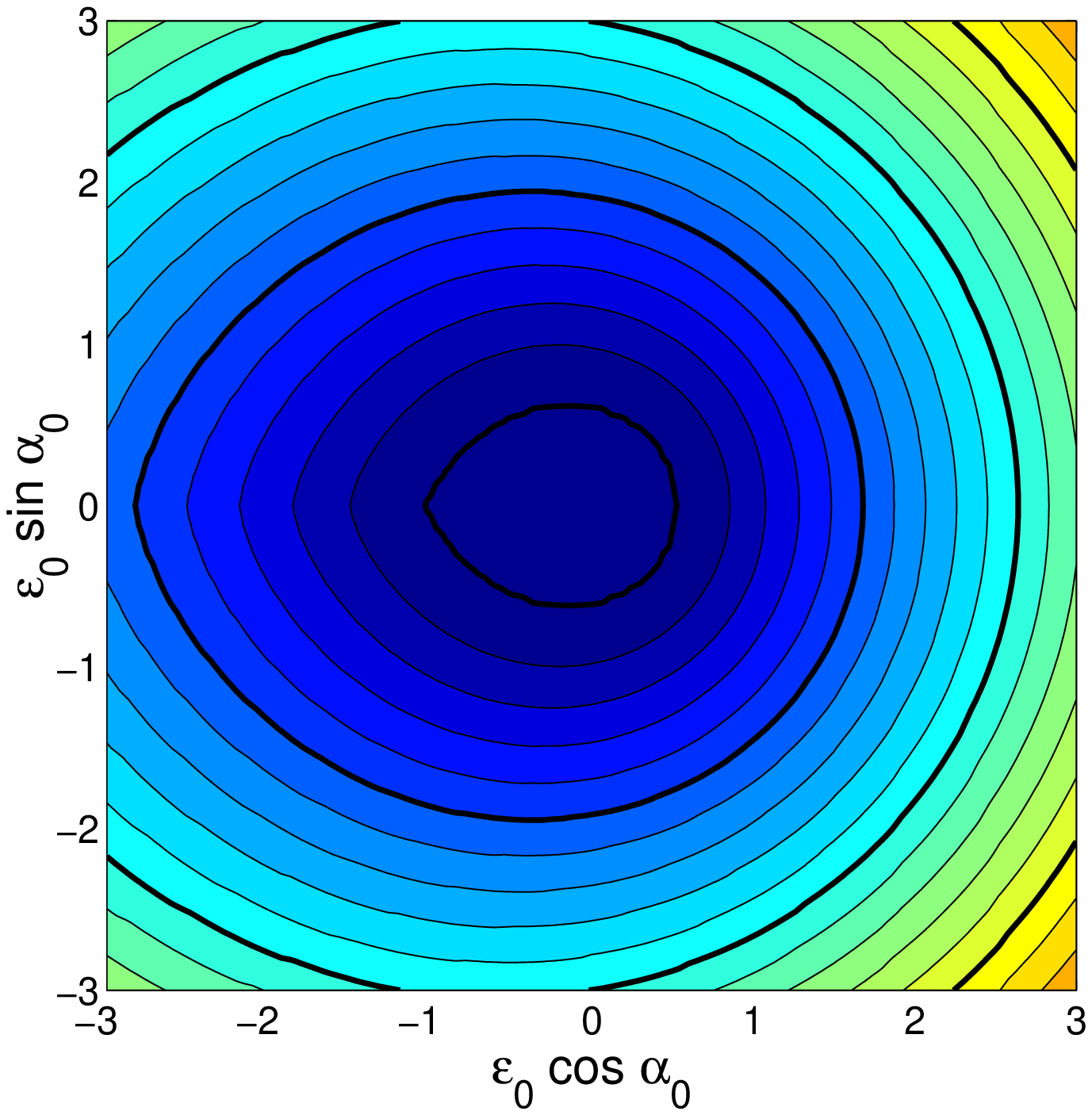}
\end{figure}

\end{document}